\documentclass[english,aps,amssymb,nofootinbib,notitlepage]{revtex4-1}
\usepackage[T1]{fontenc}
\usepackage[utf8]{inputenc}
\usepackage{babel}
\usepackage{prettyref}
\usepackage{float}
\usepackage{amsmath}
\usepackage{amssymb}
\usepackage{wasysym}
\usepackage{graphicx}
\usepackage{esint}
\usepackage[unicode=true,
 bookmarks=false,
 breaklinks=false,pdfborder={0 0 1},backref=section,colorlinks=false]
 {hyperref}
\hypersetup{
 pdfauthor={Santiago Casas},
 pdfsubject={Cosmology},
 pdfkeywords={Cosmology},
 colorlinks,linkcolor={red!50!black},citecolor={blue!50!black},urlcolor={blue!80!black}}

\makeatletter

\providecommand{\tabularnewline}{\\}

\@ifundefined{textcolor}{}
{%
 \definecolor{BLACK}{gray}{0}
 \definecolor{WHITE}{gray}{1}
 \definecolor{RED}{rgb}{1,0,0}
 \definecolor{GREEN}{rgb}{0,1,0}
 \definecolor{BLUE}{rgb}{0,0,1}
 \definecolor{CYAN}{cmyk}{1,0,0,0}
 \definecolor{MAGENTA}{cmyk}{0,1,0,0}
 \definecolor{YELLOW}{cmyk}{0,0,1,0}
}

\usepackage{babel}
\usepackage{babel}
\usepackage{babel}
\usepackage{babel}
\usepackage{babel}
\usepackage{babel}
\usepackage{babel}
\usepackage{babel}
\usepackage{babel}
\usepackage{babel}
\usepackage{morefloats}
\usepackage{xcolor}

\@ifundefined{textcolor}{}{%
 \definecolor{BLACK}{gray}{0}
 \definecolor{WHITE}{gray}{1}
 \definecolor{RED}{rgb}{1,0,0}
 \definecolor{GREEN}{rgb}{0,1,0}
 \definecolor{BLUE}{rgb}{0,0,1}
 \definecolor{blue}{rgb}{0,0,1}
 \definecolor{CYAN}{cmyk}{1,0,0,0}
 \definecolor{MAGENTA}{cmyk}{0,1,0,0}
 \definecolor{YELLOW}{cmyk}{0,0,1,0}
}

\usepackage{placeins}

\makeatother

\begin{document}

\title{Dynamics of neutrino lumps in growing neutrino quintessence}

\author{Santiago Casas$^{1}$}

\email{casas@thphys.uni-heidelberg.de}

\author{Valeria Pettorino$^{1}$}

\email{v.pettorino@thphys.uni-heidelberg.de}

\author{Christof Wetterich$^{1}$}

\email{c.wetterich@thphys.uni-heidelberg.de}

\affiliation{$^{1}$Institut f{ü}r Theoretische Physik, Ruprecht-Karls-Universit{ä}t
Heidelberg, Philosophenweg 16, 69120 Heidelberg, Germany}
\begin{abstract}
We investigate the formation and dissipation of large scale neutrino
structures in cosmologies where the time evolution of dynamical dark
energy is stopped by a growing neutrino mass. In models where the
coupling between neutrinos and dark energy grows with the value of
the scalar cosmon field, the evolution of neutrino lumps depends on
the neutrino mass. For small masses the lumps form and dissolve periodically,
leaving only a small backreaction of the neutrino structures on the
cosmic evolution. This process heats the neutrinos to temperatures
much above the photon temperature such that neutrinos acquire again
an almost relativistic equation of state. The present equation of
state of the combined cosmon-neutrino fluid is very close to -1. In
contrast, for larger neutrino masses the lumps become stable. The
highly concentrated neutrino structures entail a large backreaction
similar to the case of a constant neutrino-cosmon coupling. A present
average neutrino mass of around $0.5$ eV seems so far compatible
with observation. For masses lower than this value, neutrino induced
gravitational potentials remain small, making the lumps difficult
to detect. 
\end{abstract}
\maketitle

\section{Introduction}

\global\long\def\lcdm{\Lambda\textrm{CDM}}

The observed accelerated expansion of the Universe is currently well
described by the standard $\lcdm$ scenario, where a cosmological
constant leads the background of the Universe to a final de Sitter
state. Such a scenario, however, raises a ``coincidence\textquotedbl{}
(or ``why-now\textquotedbl{}) problem, as it is not understood why
dark energy becomes important just recently, marking the present cosmological
epoch as a special one within cosmic history. Alternative models of
dynamical dark energy or modified gravity should address the ``why
now problem'' and the associated fine tuning of parameters. At the
same time they need to explain why dark energy is almost static in
the present epoch, such that the present dark energy equation of state
$w$ is close to the observed range near $-1$.

Growing neutrino quintessence \cite{amendola_growing_2008,wetterich_growing_2007}
explains the end of a cosmological scaling solution (in which dark
energy scales as the dominant background) and the subsequent transition
to a dark energy dominated era by the growing mass of neutrinos, induced
by the change of the value of the cosmon field which is responsible
for dynamical dark energy. The dependence of the mass of neutrinos
on the cosmon (dark energy) field $\phi$, 
\begin{equation}
m_{\nu}=m_{\nu}(\phi)\propto\hat{m}_{\nu}e^{-\int\beta(\phi)d\phi}\,\,,\beta(\phi)=-\frac{\partial\ln m_{\nu}(\phi)}{\partial\phi}\label{eq: mass_def}
\end{equation}
involves the cosmon-neutrino coupling $\beta(\phi)$ which measures
the strength of the fifth force (additional to gravity). The constant
$\hat{m}_{\nu}$ is a free parameter of the model which determines
the size of the neutrino mass. (We take for simplicity all three neutrino
masses equal - or equivalently $m_{\nu}$ stands qualitatively for
the average over the neutrino species.) The special role of the neutrino
masses (as compared to quark and charged lepton masses) is motivated
at the particle physics level by the way in which neutrinos get masses
\cite{wetterich_growing_2007}. Growing neutrino quintessence with
a sufficiently large negative value of $\beta$ successfully relates
the present dark energy density and the mass of the neutrinos. The
evolution of the cosmon is effectively stopped once neutrinos become
non-relativistic. Dark energy becomes important now because neutrinos
become non-relativistic in a rather recent past, at typical redshifts
of about $z=5$ \cite{mota_neutrino_2008}. In this way, the ``why
now problem'' is resolved in terms of a ``cosmic trigger event''
induced by the change in the effective neutrino equation of state,
rather than by relying on the fine tuning of the scalar potential.
This differs from other mass varying neutrino cosmologies (usually
known as MaVaN's) \cite{brookfield_cosmology_2007,la_vacca_mass-varying_2013,bi_cosmological_2005,fardon_dark_2004,kaplan_neutrino_2004,spitzer_stability_2006,takahashi_speed_2006}.
Some of the observational consequences of those models were studied
in \cite{la_vacca_mass-varying_2013,kaplan_neutrino_2004} and more
recently a new scalar field - neutrino coupling that produces viable
cosmologies was proposed in \cite{simpson_dark_2016}. A viable cosmic
background evolution of growing neutrino quintessence offers interesting
prospects of a possible observation of the neutrino background.

The case in which the coupling $\beta$ is constant has been largely
investigated in literature at the linear level \cite{mota_neutrino_2008},
in semi-analytical non-linear methods \cite{wintergerst_clarifying_2010,wintergerst_very_2010,brouzakis_nonlinear_2011},
joining linear and non-linear information to test the effect of the
neutrino lumps on the cosmic microwave background \cite{pettorino_neutrino_2010}
and within N-Body simulations \cite{ayaita_neutrino_2013,ayaita_structure_2012,baldi_oscillating_2011,ayaita_nonlinear_2016}.
For the values of $\beta$ ($\beta\apprge10^{3}$) needed for dark
energy to dominate today, the cosmic neutrino background is clumping
very fast. Large and concentrated neutrino lumps form and induce very
substantial backreaction effects. These effects are so strong that
the deceleration of the evolution of the cosmon gets too weak, making
it difficult to obtain a realistic cosmology \cite{fuhrer_backreaction_2015}.

In this paper we instead consider the case in which the neutrino-cosmon
coupling $\beta(\phi)$ depends on the value of the cosmon field and
increases with time. In a particle physics context this has been motivated
\cite{wetterich_growing_2007} by a decrease with $\phi$ of the heavy
mass scale (B-L-violating scale) entering inversely the light neutrino
masses. In this scenario $\beta(\phi)$ has not been large in all
cosmological epochs - the present epoch corresponds to a crossover
where $\beta$ gets large. A numerical investigation \cite{baldi_oscillating_2011}
of this type of model has revealed compatibility with observations
for the case of a present neutrino mass $m_{\nu,0}=0.07$ eV. In the
present paper we investigate the dependence of cosmology on the value
of the neutrino mass by varying the parameter $\hat{m}_{\nu}$ in
eq. \prettyref{eq: mass_def}. For large neutrino masses we find a
qualitative behavior similar to the case of a constant neutrino-cosmon
coupling $\beta$, with difficulties to obtain a realistic cosmology.
In contrast, for small neutrino mass, the neutrino lumps form and
dissolve, with small influence on the overall cosmological evolution.
In this case, the neutrino-induced gravitational potentials are found
to be much smaller than the ones induced by dark matter. As we will
discuss in this paper, it will not be easy to find observational signals
for the neutrino lumps. In-between the regions of small and large
neutrino masses we expect a transition region for intermediate neutrino
masses where, by continuity, observable effects of the neutrino lumps
should show up.

\section{Growing neutrinos with varying coupling}

We consider here cosmologies in which neutrinos have a mass that varies
in time, along the framework of ``varying growing neutrino models''
\cite{wetterich_growing_2007}. As long as neutrinos are relativistic,
the coupling is inefficient and the dark energy scalar field $\phi$
rolls down a potential, as in an early dark energy scenario. As the
neutrino mass increases with time, neutrinos become non-relativistic,
typically at a relatively late redshift $z\approx4-6$ \cite{pettorino_neutrino_2010}.
This influences the evolution of $\phi$, which feels the effect of
neutrinos via a coupling to the neutrino mass $m_{\nu}(\phi)$. The
evolution of the scalar field slows down and practically stops, such
that the potential energy of the cosmon behaves almost as a cosmological
constant at recent times. In other words, in these models the cosmological
constant behavior observed today is related to a cosmological trigger
event (i.e. neutrinos becoming non-relativistic) and the present dark
energy density is directly connected to the value of the neutrino
mass. In the following we will detail the formalism and equations
used to describe the cosmological evolution of the model.

We start with the linearized Friedman-Lemaitre-Robertson-Walker metric
in the Newtonian gauge: 
\begin{equation}
ds^{2}=-(1+2\Psi)dt^{2}+a^{2}(1-2\Phi)d\mathbf{x}^{2}\,\,.
\end{equation}
Moreover, we use a quasi-static approximation for sub-horizon scales
($H/k\ll1$) which allows us to neglect time derivatives with respect
to spatial ones. Then the quasi-static, first-order perturbed Einstein
equations, are the Poisson equation \cite{ma_cosmological_1994}

\begin{equation}
k^{2}\Phi=4\pi Ga^{2}\delta T_{0}^{0}\,\,,\label{eq:Poisson}
\end{equation}
and the ``stress'' equation 
\begin{equation}
k^{2}(\Phi-\Psi)=12Ga^{2}(\bar{\rho}+\bar{P})\sigma\,\,,\label{eq:stress-phipsi}
\end{equation}
where $\delta T_{0}^{0}$ is the perturbation of the $0-0$ component
of the energy momentum tensor $T_{\mu\nu}$ and $\sigma$ is the anisotropic
stress of the fluid which depends on the traceless component of the
spatial part of the energy-momentum tensor, $T_{j}^{i}-\delta_{j}^{i}T_{k}^{k}/3$.
This stress tensor is in our case only important for relativistic
particles (i.e. the neutrinos). The source term of the Poisson equation
\prettyref{eq:Poisson} will contain contributions from all matter
species (dark matter \& neutrinos) and from the cosmon field. It is
proportional to the total density contrast $\delta\rho_{t}=\delta\rho_{\nu}+\delta\rho_{m}+\delta\rho_{\phi}$.

The cosmon field can be described through a Lagrangian in the standard
way 
\begin{equation}
-\mathcal{L_{\phi}}=\frac{1}{2}\partial^{\nu}\phi\partial_{\nu}\phi+V(\phi)
\end{equation}
where for this work we choose an exponential potential $V(\phi)\propto e^{-\alpha\phi}$.
The field dependent mass (eq.\prettyref{eq: mass_def}) allows for
an energy-momentum transfer between neutrinos and the cosmon, which
is proportional to the trace of the energy momentum tensor of neutrinos
$T_{(\nu)}$ and to a coupling parameter $\beta(\phi)$ 
\begin{align}
\nabla_{\eta}T_{(\phi)}^{\mu\eta}= & +\beta(\phi)T_{(\nu)}\partial^{\mu}\phi\,\,\,,\label{eq:continuity-phi}\\
\nabla_{\eta}T_{(\nu)}^{\mu\eta}= & -\beta(\phi)T_{(\nu)}\partial^{\mu}\phi\,\,\,.\label{eq:continuity-nu}
\end{align}

The cosmon is the mediator of a fifth force between neutrinos, acting
at cosmological scales. Its evolution is described by the Klein-Gordon
equation sourced by the trace of the energy-momentum tensor $T_{(\nu)}$
of the neutrinos,

\begin{equation}
\nabla_{\mu}\nabla^{\mu}\phi-V'(\phi)=\beta(\phi)T_{(\nu)}.\label{eq:klein-gordon-equation}
\end{equation}
As long as the neutrinos are relativistic ($T_{(\nu)}=0)$ the source
on the right hand side vanishes. During this time, the coupling has
no effect on the evolution of $\phi$. While the potential term $\sim V'$
drives $\phi$ towards larger values, the term $\sim\beta$ has the
opposite sign and stops the evolution effectively once $\beta T_{(\nu)}$
equals $V'$. The trace of the energy momentum tensor $T_{\nu}$,
entering eq.\prettyref{eq:klein-gordon-equation} is equal to: 
\begin{equation}
T_{\nu}=m_{\nu}(\phi)\tilde{n}(\phi)\label{eq:trace}
\end{equation}
where $\tilde{n}_{\nu}(\phi)=n_{\nu}(\phi)/\gamma$ is the ratio of
the number density of neutrinos $n_{\nu}$, divided by the relativistic
$\gamma$ factor. Eq.\prettyref{eq:trace} is valid for both relativistic
and non-relativistic neutrinos. Here we consider a coupling $\beta$
between neutrino particles and the quintessence scalar field $\phi$
as a field dependent quantity: 
\begin{equation}
\beta(\phi)\equiv-\frac{1}{\phi_{c}-\phi}\,.\label{eq:beta-of-phi}
\end{equation}
From eq.\prettyref{eq: mass_def} the neutrino mass is then given
by: 
\begin{equation}
m_{\nu}(\phi)=\frac{\bar{m}_{\nu}}{\phi_{c}-\phi}\,.\label{eq:mnu-of-phi}
\end{equation}
Here $\phi_{c}$ denotes the asymptotic value of $\phi$ for which
$\beta$ and $m_{\nu}(\phi)$ would formally become infinite. By an
additive shift in $\phi$ it can be set to an arbitrary value, e.g.
$\phi_{c}=0$. We consider the range $\phi<\phi_{c}$. The divergence
of $\beta$ for $\phi\rightarrow\phi_{c}$ in eq.\prettyref{eq:beta-of-phi}
is not crucial for the results of the present paper - $\beta$ and
$m_{\nu}$ never increase to large values, such that the immediate
vicinity of $\phi_{c}$ plays no role.

The coupling induces a total force acting on neutrinos given by $\nabla(\Phi_{\nu}+\beta\delta\phi)$
and appearing in the corresponding Euler equation \cite{pettorino_neutrino_2010},
as usual in coupled cosmologies \cite{baldi_hydrodynamical_2010}.
For values $2\beta^{2}>1$ the fifth force induced on neutrinos by
the cosmon becomes larger than the gravitational attraction. For the
large values of $|\beta|\approx10^{2}$ reached during the cosmological
evolution, the attraction induced by the cosmon gives rise to the
formation of neutrino lumps. As shown in \cite{mota_neutrino_2008,pettorino_neutrino_2010}
this represents the major difficulty encountered within growing neutrino
models and also, simultaneously, one of its clearest predictions with
respect to alternative dark energy models: the presence of neutrino
lumps at scales of $\approx10$ Mpc or even larger, depending on the
details of the model \cite{mota_neutrino_2008}. Since the attractive
force between neutrinos is $10^{4}$ times bigger than gravity, therefore
also the dynamical time scale of the clumping of neutrino inhomogeneities
is a factor $10^{4}$ faster than the gravitational time scale. Even
the tiny inhomogeneities in the cosmic neutrino background grow very
rapidly non-linear. The impact of such structures, has been shown
to depend crucially on the strength of backreaction effects \cite{ayaita_structure_2012,ayaita_nonlinear_2016}.
For constant coupling, the effect of backreaction is strong and can
lead to neutrino lumps with rapidly growing concentration, reaching
values of the gravitational potential which exceed observational constraints.
The effect is so strong that it is able to destroy the oscillatory
effect first encountered in \cite{baldi_hydrodynamical_2010}, in
which neutrino lumps were forming and then dissipating. No realistic
cosmology has been found in this case \cite{fuhrer_backreaction_2015}.
With the varying coupling of eq.\prettyref{eq:beta-of-phi} a similar
behavior will be found for large neutrino masses. For small neutrino
masses the oscillatory effects will be dominant and realistic cosmologies
seem possible \cite{ayaita_nonlinear_2016}.

\section{Numerical treatment of growing neutrino cosmologies}

\subsection{Modified Boltzmann code\label{sub:Modified-Boltzmann-code}}

For the early stages of the evolution of the growing neutrino quintessence
model, neutrinos behave as standard relativistic particles and the
coupling to the cosmon field is suppressed. Therefore the Klein-Gordon
equation can be linearized and no important backreaction effects are
present. The Einstein-Boltzmann system of equations for the relativistic
neutrinos and all other species has been solved using a modified version
of the code CAMB \cite{lewis_efficient_2000} (hereafter referred
to as nuCAMB), used and developed already in previous papers on mass-varying
and growing neutrino cosmologies. We refer the reader to previous
publications \cite{mota_neutrino_2008,pettorino_neutrino_2010,wintergerst_very_2010,brookfield_cosmology_2007}
for details about its implementation. These equations are valid until
neutrinos become non-relativistic and as long as perturbations are
still linear. The neutrinos can be seen as a weakly-interacting gas
of particles in thermal equilibrium with a phase space distribution
$f(p)$ with $p$ denoting the momentum. The statistical description
is in the case of neutrinos a Fermi-Dirac distribution, given by 
\begin{equation}
f_{FD}(p)=\frac{1}{e^{\left(E(p)-\mu\right)/T}+1}\,\,,\label{eq:Fermi-Dirac-dist}
\end{equation}
where $\mu$ is the chemical potential and $E(p)=\sqrt{m^{2}+p^{2}}$
the particle energy. Then, the number density of neutrinos, the energy
density and the pressure are given respectively by

\begin{align}
n_{\nu} & =\frac{2}{(2\pi)^{3}}\int\mbox{d}^{3}p\, f_{FD}(p)\,\,,\label{eq:nnu-FD}\\
\rho_{\nu} & =\frac{2}{(2\pi)^{3}}\int\mbox{d}^{3}p\, E(p)f_{FD}(p)\,\,,\label{eq:rhonu-FD}\\
P_{\nu} & =\frac{2}{(2\pi)^{3}}\int\mbox{d}^{3}p\,\frac{p^{2}}{E(p)}f_{FD}(p)\,\,.\label{eq:pnu-FD}
\end{align}
The solution of the Boltzmann hierarchy of neutrinos coupled to the
perturbed Einstein equations \prettyref{eq:Poisson} and \prettyref{eq:stress-phipsi},
together with the solution of the background Klein-Gordon equation
\prettyref{eq:klein-gordon-equation}, form the basis of the modification
of nuCAMB with respect to the standard code CAMB, which handles dark
matter, photons and baryons altogether. We recall that for growing
neutrino quintessence, the neutrino mass depends on the cosmon field
$\phi$ and therefore on the scale factor $a$.

The ratio of the initial mass of the neutrinos to their temperature
(given in eV), is calculated in nuCAMB as follows 
\begin{equation}
\hat{r}_{\nu eV}\equiv\left(\frac{m}{T}\right)_{\nu,camb}=\frac{(7/8)(\pi^{4}/15)}{(3/2)\zeta(3)}\times\frac{\rho_{cr}\Omega_{\nu,input}}{\text{\ensuremath{\rho}}_{\nu}}\,\,.
\end{equation}
The first fraction comes from the relation $m_{\nu}\approx\text{\ensuremath{\rho}}_{\nu}/n_{\nu}=((\frac{7}{8}\frac{\pi^{4}}{15})/\frac{3}{2}\zeta(3))T_{\nu}$,
which is valid in the non-relativistic limit of eqns.\prettyref{eq:Fermi-Dirac-dist}-\prettyref{eq:pnu-FD};
the critical density is defined as usual: $\rho_{cr}=\frac{3H_{0,input}^{2}}{8\pi G}$.
The second fraction is a re-scaling that corrects the neutrino density
in order to match the wanted $\Omega_{\nu,input}$ given as input
value. The code performs an iterative routine that varies initial
conditions in such a way that the input parameters are obtained at
present time. Since this is not exact, the final values of $H_{0}$
and $\Omega_{\nu}$ might vary slightly with respect to the given
input values. The ratio $\hat{r}_{\nu eV}$ depends on the input parameters
$H_{0,input}$ (via the critical density) and on $\Omega_{\nu,input}$
\footnote{$\hat{r}_{\nu eV}$ is also the conversion factor between the mass
units in the N-Body code and units in eV.%
}. Furthermore, the neutrino energy density $\rho_{\nu}$ and the photon
energy density $\rho_{\gamma}$ at relativistic times are related
as 
\begin{equation}
\rho_{\nu}=N_{\nu}\times\frac{7}{8}\times\left(\frac{4}{11}\right)^{4/3}\rho_{\gamma}
\end{equation}
where $N_{\nu}=3$ is the number of neutrino species. The use of these
formulae is valid if initial conditions are set when neutrinos are
non-relativistic, where the linear regime still applies. For initial
conditions set at earlier time relativistic corrections have to be
taken into account. After solving the Einstein-Boltzmann system, realizations
of the fields $\delta_{\nu}(\mathbf{k})$ and $v_{pec,\nu}(\mathbf{k})$
at an early time are obtained from nuCAMB and are then used as the
initial conditions for the neutrino distribution in the growing neutrino
quintessence N-body simulation. This will be explained more in detail
at the end of the following section.

\subsection{N-body simulation\label{sub:N-body-simulation}}

For N-body simulations, we use here the code developed in \cite{ayaita_neutrino_2013,ayaita_nonlinear_2016,ayaita_structure_2012}
and then refined in \cite{fuhrer_backreaction_2015} and in the present
work, which uses a particle-mesh approach for the neutrino and dark
matter particle evolution and a multi-grid approach for solving the
non-linear scalar field equations. In table \ref{tab:simus-params}
we describe the parameters of the models discussed in this paper.
We consider 5 models with different neutrino masses.

Our N-body simulation differs from standard Newtonian N-body codes
in many ways, the most important one being that we evolve the cosmon
$\phi$ and the gravitational potentials $\Phi$ and $\Psi$ separately.
While neutrinos, dark matter and the cosmon are non-linear in the
N-body simulations, we assume that the gravitational potentials $\Phi$
and $\Psi$ are small, which is valid in cosmological applications,
even for large deviations of standard $\lcdm$ and at small scales.
The perturbation in the dark energy scalar field $\delta\rho_{\phi}$
can be calculated from the perturbation of the energy density of the
cosmon field 
\begin{equation}
\delta\rho_{\phi}=\frac{\bar{\phi}'\delta\phi}{a^{2}}+V(\bar{\phi})\delta\phi\,\,\,.\label{eq:perturbed-cosmon}
\end{equation}
The evolution of the homogeneous potential of the cosmon field can
be obtained through its energy density and pressure in the following
way 
\begin{equation}
V_{\phi}(a)=\frac{1}{2}(\rho_{\phi}(a)-p_{\phi}(a))\,\,,
\end{equation}
while the perturbations in the potential can be approximated by 
\begin{equation}
\delta V_{\phi}(a)=-\frac{1}{2}(\delta\rho_{\phi}(a)+3\delta p_{\phi}(a))\,\,.
\end{equation}
The cosmon field can cluster and therefore its spatial gradients are
non-vanishing, so that after averaging over the volume of the box
the energy density of the cosmon field is 
\begin{equation}
\bar{\rho}_{\phi}=\frac{1}{2}\overline{\dot{\phi}^{2}}+\frac{1}{2a^{2}}\overline{(1+2\Phi)\partial_{i}\phi\partial_{j}\phi\delta^{ij}}+\overline{V(\phi)}\,\,,
\end{equation}
while its pressure reads 
\begin{equation}
\bar{P}_{\phi}=\frac{1}{2}\overline{\dot{\phi}^{2}}-\frac{1}{6a^{2}}\overline{(1+2\Phi)\partial_{i}\phi\partial_{j}\phi\delta^{ij}}+\overline{V(\phi)}\,\,.
\end{equation}
We will use for the following a convention in which bars denote spatial
averages, while angular brackets denote time averaged quantities.
The evolution of the cosmon field is solved using a multigrid relaxation
algorithm, known as the Newton-Gauß-Seidel solver, which was originally
developed for $f(R)$ modified gravity simulations \cite{puchwein_modified-gravity-gadget:_2013}
and has also been implemented into the growing neutrino N-body simulations
in \cite{ayaita_nonlinear_2016}. The bottom part of table \ref{tab:simus-params}
lists the results of the six models computed using the N-body simulations.

In the case of neutrinos, the mass is a time-varying quantity following
eq.\prettyref{eq:mnu-of-phi}. Neutrinos obey a modified geodesic
equation 
\begin{equation}
\frac{du^{\mu}}{d\tau}+\Gamma_{\nu\lambda}^{\mu}u^{\nu}u^{\lambda}=\beta(\phi)\partial^{\mu}\phi+\beta(\phi)u^{\nu}u^{\mu}\partial_{\nu}\phi\,\,,
\end{equation}
in which the right hand side gets a contribution from the coupling.

\begin{table}
\centering{}%
\begin{tabular}{|c|c|c|c|c|c|c|}
\hline 
\textbf{Cosmological parameters}  & \multicolumn{6}{c|}{\textbf{Growing neutrino models}}\tabularnewline
\hline 
\textbf{{Linear values} }  & \textbf{{M1} }  & \textbf{{M2} }  & \textbf{{M3} }  & \textbf{{M4} }  & \textbf{{M5} }  & \textbf{{M6}}\tabularnewline
\hline 
\hline 
$\Omega_{\nu0}+\Omega_{\phi0}$  & 0.686  & 0.688  & 0.692  & 0.701  & 0.693  & 0.697\tabularnewline
\hline 
$\Omega_{\nu0}$  & $3.8\times10^{-3}$  & $2.6\times10^{-2}$  & $1.64\times10^{-2}$  & $4.7\times10^{-2}$  & $6.1\times10^{-2}$  & $9.4\times10^{-2}$\tabularnewline
\hline 
$h$  & 0.671  & 0.673  & 0.6818  & 0.701  & 0.722  & 0.740\tabularnewline
\hline 
$m_{\nu0}$ {[}eV{]}  & $0.060$  & $0.407$  & $0.239$  & $0.730$  & $1.000$  & $1.712$\tabularnewline
\hline 
$\left\langle m_{\nu}\right\rangle $$[0.4:0.6]${[}eV{]}  & $0.040$  & $0.067$  & $0.134$  & $0.277$  & $0.399$  & $0.701$\tabularnewline
\hline 
$\left\langle m_{\nu}\right\rangle $$[0.8:1.0]${[}eV{]}  & $0.099$  & $0.152$  & $0.318$  & $0.661$  & $0.907$  & $1.51$\tabularnewline
\hline 
$\langle w_{\nu\phi}\rangle[0.9:1.0]$  & -0.97  & -0.97  & -0.95  & -0.92  & -0.90  & -0.85\tabularnewline
\hline 
\hline 
\textbf{{N-body values} }  &  &  &  &  &  & \tabularnewline
\hline 
\hline 
$\Omega_{\nu0}+\Omega_{\phi0}$  & 0.688  & 0.690  & -  & -  & -  & -\tabularnewline
\hline 
$\Omega_{\nu0}$  & $2.5\times10^{-2}$  & $1.9\times10^{-2}$  & -  & -  & -  & -\tabularnewline
\hline 
$m_{\nu0}$ {[}eV{]}  & 0.038  & 0.078  & -  & -  & -  & -\tabularnewline
\hline 
$\left\langle m_{\nu}\right\rangle $$[0.4:0.6]$ {[}eV{]}  & $0.048$  & $0.069$  & $0.1436$  & $0.280$  & $0.401$  & $0.676$\tabularnewline
\hline 
$\left\langle m_{\nu}\right\rangle [0.8:1.0]$ {[}eV{]}  & $0.120$  & $0.164$  & -  & -  & -  & -\tabularnewline
\hline 
$\langle w_{\nu\phi}\rangle[0.9:1.0]$  & -0.95  & -0.96  & -  & -  & -  & -\tabularnewline
\hline 
$a_{\text{final}}$  & 1.0  & 1.0  & 0.84  & 0.70  & 0.65  & 0.67\tabularnewline
\hline 
\end{tabular}\protect\protect\protect\caption{\label{tab:simus-params} Table of parameters for the six models considered
in this work. The top part refers to the output values computed with
the linear nuCAMB code. The bottom part refers to values computed
within the N-body simulation. Quantities denoted with a subscript
$0$ are values at present time, $a=1.0.$ The $\left\langle m_{\nu}\right\rangle [a_{1}:a_{2}]$
is the root mean squared (RMS) value of the neutrino mass in units
of eV computed between $a=a_{1}$ and $a=a_{2}$. The same notation
is also valid for $\langle w_{\nu\phi}\rangle[a_{1},:a_{2}]$ corresponding
to the equation of state of the combined cosmon and neutrino fluid
which represents dynamical dark energy. $a_{final}$ is the final
time at which simulations were computed accurately. Therefore, for
the models M3-M6 we cannot cite values of present time quantities
or averages at times beyond $a_{final}$. The input values for nuCAMB
corresponding to all models can be found in table \ref{tab:CAMB-inipars-1}
of appendix \ref{sec:Initial-parameters-for-nucamb}.}
\end{table}

Simulations start at an initial value of $a_{in}=0.02$. Until $a\approx0.30$
the dark matter particles, the cosmon field and the gravitational
potentials are evolved on the grid. For dark matter particles we take
standard initial conditions from nuCAMB and start the particle-mesh
algorithm that solves the Poisson equation \prettyref{eq:Poisson}
at an initial redshift of $z=49$. This is not the most accurate way
of setting initial conditions for cosmological dark matter simulations
(see for example recent N-body comparisons by \cite{schneider_matter_2016}),
but since in this work we are not interested in detailed substructures
of dark matter halos or a percent-accurate power spectrum, we find
that our approach gives a correct description at the scales of interest.
Neutrinos are first treated differently from other particles, as a
distribution of relativistic particles in thermal equilibrium and
no backreaction effects from neutrino structures are taken into account.
Starting from a scale factor of approximately $a_{ini}\approx0.30$
(depending on the exact parameters of each model), which is when neutrinos
become non-relativistic, neutrinos are also projected on the grid:
their phase-space distribution is sampled using effective particles.
Since their equation of state is non-relativistic, we can approximate
the phase-space distribution by 
\begin{equation}
f_{\nu}(\mathbf{x},\mathbf{v})=\bar{n}_{\nu}f_{FD}(|\mathbf{v}_{\nu}-\mathbf{v_{pec\mathbf{,}\nu}}(\mathbf{x})|)(1+\delta_{v}(\mathbf{x}))\,\,\,,\label{eq:thermal-velocities}
\end{equation}
where $f_{FD}$ is the Fermi-Dirac distribution (\prettyref{eq:Fermi-Dirac-dist}).
The thermal velocities of the neutrinos are the difference between
their total velocities and their peculiar velocities $\mathbf{v}_{th,\nu}=\mathbf{v_{\nu}}-\mathbf{v}_{pec,\nu}$.
We obtain $\delta_{v}(\mathbf{x})$ and $\mathbf{v_{pec\mathbf{,}\nu}}(\mathbf{x})$
by Fourier transforming the momentum-space realization of those fields
obtained at the time $a_{ini}$ from nuCAMB. Equation \prettyref{eq:thermal-velocities}
is solved for $\mathbf{v}_{th,\nu}$ in order to obtain the correct
thermal distribution of particles and we duplicate the number of neutrino
particles in each grid, assigning to each of them a thermal velocity
which is equal in magnitude but opposite in direction, to avoid a
distortion of the distribution of peculiar velocities at larger scales
than a single grid cell size. For a large enough number of effective
neutrino particles (i.e. when there is much more than one particle
per cell), the distortion of the peculiar velocities by thermal velocities
should be negligible. The correct neutrino density one would obtain
from the Fermi-Dirac distribution for a non-relativistic particle
reads 
\begin{equation}
\left\langle \rho_{\nu}(\mathbf{x})\right\rangle _{f_{\nu}}=\int\mbox{d}^{3}v\: m_{\nu}f_{\nu}(\mathbf{x},\mathbf{v})=m_{\nu}\bar{n}_{\nu}(1+\delta_{v}(\mathbf{x}))\,\,.\label{eq:rhoFD-nr}
\end{equation}
Since we need to enforce the right hand side of \prettyref{eq:rhoFD-nr}
at each grid cell of comoving volume $a^{3}\Delta V$, where the mass
of the neutrinos is given by the scalar field, we have a condition
on the number of particles $N_{part}$, such that 
\begin{equation}
\frac{M_{\nu}\left\langle N_{part}\right\rangle }{a^{3}\Delta V}=m_{\nu}\bar{n}_{\nu}(1+\delta_{v}(\mathbf{x}))\,\,,
\end{equation}
is fulfilled (more details of these method can be found in \cite{ayaita_structure_2012}).
When neutrinos enter as particles into the N-body simulation and therefore
backreaction effects from neutrino structures start becoming important,
the calculation of the fields and the potentials becomes computationally
demanding, due to the non-linearity of the terms sourcing the continuity
\prettyref{eq:continuity-nu} and Klein-Gordon equations \prettyref{eq:klein-gordon-equation}.
Since these equations cannot be linearized due to the large values
of the coupling parameter $\beta(\phi)$, the multigrid Newton-Gauß-Seidel
solver is of crucial importance. For the parallelization of the code,
we use a simple \emph{OpenMP} approach, which calculates in parallel,
for the available processing cores, the equations of motion of the
particles and the fast Fourier transforms. In Tab.\ref{tab:Nbody-inipars-1-1}
we describe all parameters related to the N-body simulations, including
box and grid size.

\section{Lump dynamics and the low mass - high mass divide}

\begin{figure}[H]
\includegraphics[width=0.7\textheight]{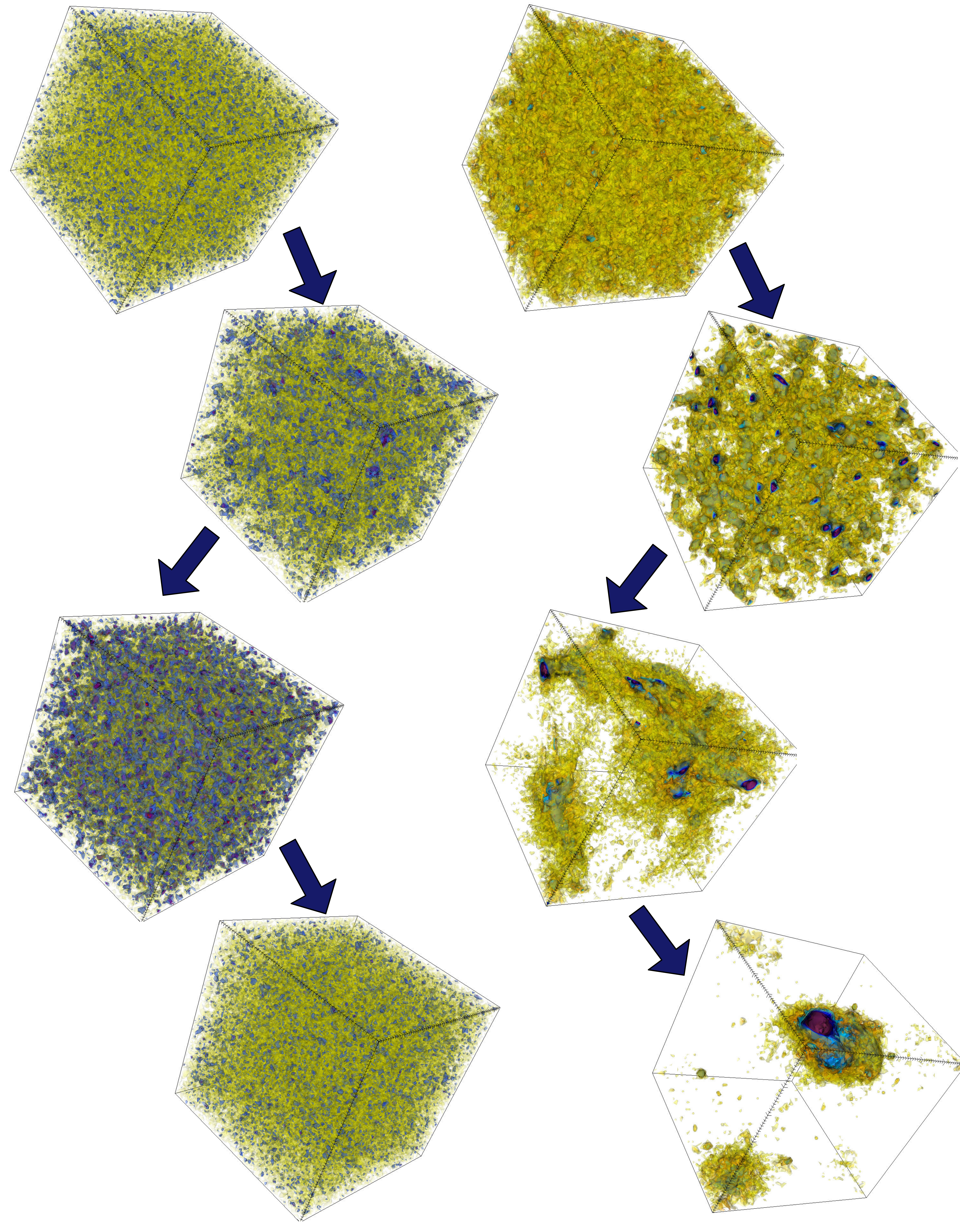}
\protect\protect\protect\caption{\label{fig:Snapshots-of-models-oscill}Snapshots of the number density
contrast\textbf{ }of neutrinos $\delta n_{\nu}(\vec{x})\equiv n_{\nu}(\vec{x})/\bar{n}_{\nu}-1$
at different times. \textbf{Left:} Model M2, at scale factors $a=0.45,\,0.7,\,0.75$
and $0.95$ from top to bottom. The overdensity oscillates between
values close to 1 (represented as yellow tones) at early times, where
there are no lumps, to values close to 10 (dark blue and purple tones),
where several concentrated lumps form at intermediate times. At later
times lumps dissolve and the overdensity decreases back to values
close to unity. \textbf{Right: }Model M4, at scale factors: $a=0.35,\,0.42,\,0.53$
and $0.64$ from top to bottom. The neutrino lumps start growing at
early times and merge progressively into larger and more concentrated
structures. At the end, almost all neutrinos are attracted to a single
very massive lump.}
\end{figure}

We find two different regimes for the non-linear evolution of neutrino
lumps, depending on the average value of the neutrino mass. For light
neutrino masses, during the lump formation process, the neutrinos
are accelerated to relativistic velocities. Subsequently, the lumps
dissolve and form again periodically, as described in detail in ref.
\cite{ayaita_nonlinear_2016,baldi_oscillating_2011}. We demonstrate
this behavior in the left panel of fig. \ref{fig:Snapshots-of-models-oscill}.
The repeated acceleration epochs heat the neutrino fluid to a huge
effective temperature, such that neutrinos have again an almost relativistic
equation of state during alternating periods of time.

In contrast, the behavior for large neutrino masses is qualitatively
different. The concentration of the lumps continues to grow after
their first formation. Lumps merge, and typically do not dissolve.
The neutrino number density contrast reaches high values at late times.
This is demonstrated on the right panel in fig.\ref{fig:Snapshots-of-models-oscill}
for an average value of the neutrino mass $m_{\nu,av}=0.4$eV in the
range $0.4<a<0.6$. This behavior resembles the one found for a constant
cosmon-neutrino coupling in \cite{ayaita_neutrino_2013,ayaita_structure_2012,baldi_oscillating_2011}.

\begin{figure}[H]
\begin{centering}
\includegraphics[width=0.35\textwidth]{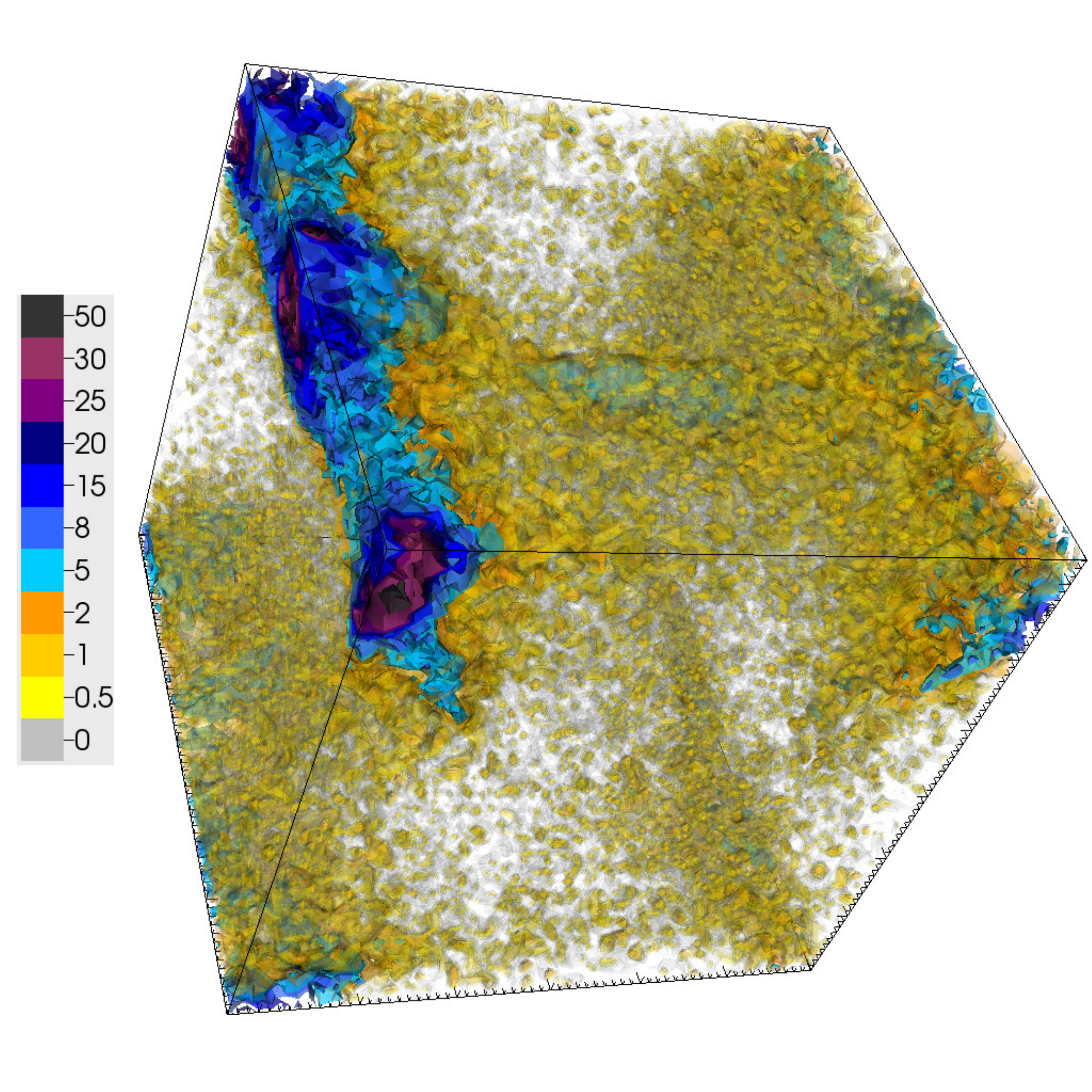}\includegraphics[width=0.35\textwidth]{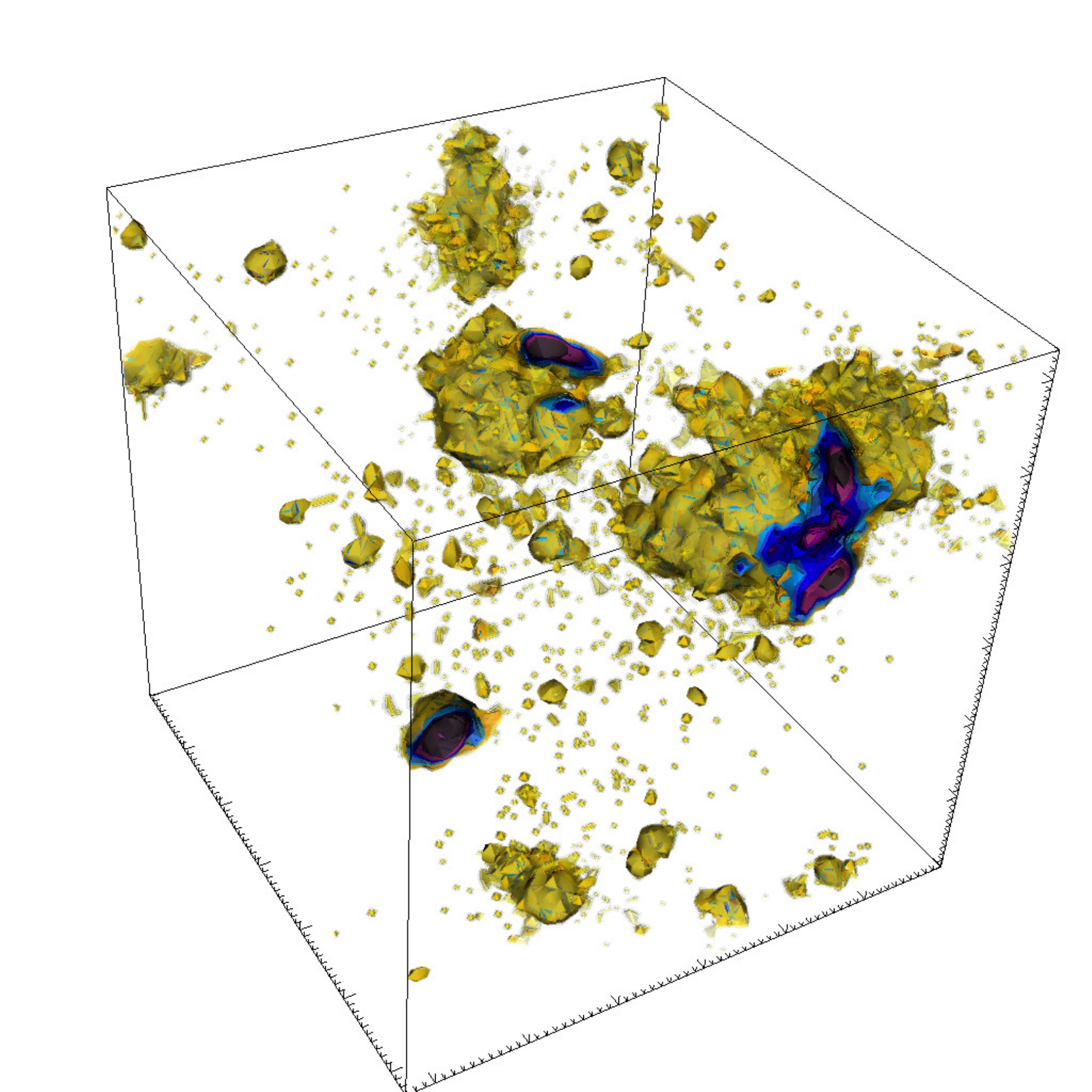} 
\par\end{centering}

\protect\protect\protect\protect\protect\protect\protect\caption{\label{fig:Snapshots-of-models-crashing}Snapshots of the neutrino
overdensity field $\delta_{\nu}(\vec{x})$ for models M4 (top left)
and M6 (bottom left) at scale factors of $a=0.64,\,$ and $0.62$,
respectively. In these models, neutrino lumps cluster into large stable
structures with a high concentration, starting from a bottom-up approach,
as was shown for model M4 in fig. \ref{fig:Snapshots-of-models-oscill}.
Neutrino structures occupy large parts of the simulation box, corresponding
to scales of \textasciitilde{}50 Mpc. At these scale factors, the
forces introduced by the cosmon coupling are too strong to be resolved
by our numerical approach and our simulation breaks down.}
\end{figure}

Due to the increasing value of the concentration and the increasing
cosmon-neutrino coupling, the characteristic time scale becomes very
short and gradients very large. This exceeds the present numerical
capability of our simulations, typically at a value of the scale factor
somewhat larger than $a=0.6$. In fig.\ref{fig:Snapshots-of-models-crashing}
we show snapshots for two different values of neutrino masses shortly
before the simulation breaks down.

The transition between the ``heating regime'' for small neutrino
masses and the ``concentration regime'' for large neutrino masses
occurs in the range $\langle m_{\nu}\rangle[0.4:0.6]\approx0.07\mbox{eV}-0.14\mbox{eV}$,
where the time average is taken for $0.4<a<0.6$. The present value
of the neutrino mass can be substantially larger due to oscillations
and the continued increase of the mass and the temperature. For example,
the phenomenologically viable model with $\langle m_{\nu}\rangle[0.4:0.6]=0.07$eV
corresponds to a present neutrino mass of around $0.08$eV, but the
time oscillations grow the neutrino mass to values of up to $0.5$eV
for very short intervals in the scale factor $a$. (compare with fig.
\ref{fig:NuMass-mod2} below).

In fig.\ref{fig:Snapshots-of-models-oscill} we show the distribution
of the number (over)density contrast $\delta n_{\nu}(\vec{x})\equiv n_{\nu}(\vec{x})/\bar{n}_{\nu}-1$
at four different times and for two different models considered here,
namely M2 (left panels) and M5 (right panels). For M2 as well as for
models with smaller masses (not shown here), the neutrino lumps form
and dissolve very quickly. The lumps are never stable and neutrinos
accelerate to relativistic velocities when they fall into the gravitational
potentials. The small lumps are also distributed homogeneously across
the simulation box (see the third panel from above on the right of
fig.\ref{fig:Snapshots-of-models-oscill}). The lumps reach maximal
number density contrasts of about $\delta n_{\nu}\approx10$. For
M5 and for bigger masses, the neutrino lumps become stable, accreting
more and more particles with the passing of time and increasing their
concentrations. This leads to strong backreaction effects, changing
the background cosmological evolution. After some time all neutrinos
are concentrated in very big lumps, reaching very high values of $\delta{}_{\nu}\approx50-100$,
where $\delta_{\nu}\equiv\rho_{\nu}(\vec{x})/\bar{\rho}_{\nu}-1$,
see fig.\ref{fig:Snapshots-of-models-crashing}. After this point,
the numerical framework for the growing neutrino quintessence evolution
breaks down and we can no longer solve reliably the coupled system
of equations.

\section{Cosmological evolution in the light neutrino regime}

As we have seen in the previous section, there is a qualitative difference
between the cosmological evolution of a model with a light or a heavy
neutrino mass, the boundary being a present neutrino mass value of
roughly $\approx0.5$eV (calculated in linear theory). In this section
we explore more in detail the evolution of background quantities in
the light mass model M2 whose parameters are shown in detail in table
\ref{tab:simus-params} for the linear calculation in nuCAMB (top
panel) and for the N-body computation (bottom panel). We study in
detail the differences appearing in the evolution of background quantities,
when non-linear physics and backreaction are taken into account.

The standard definition for the homogeneous energy density fraction
of the cosmon field $\phi$ is 
\begin{equation}
\Omega_{\phi}=\frac{8\pi G}{3H^{2}}\bar{\rho}_{\phi}\,\,,
\end{equation}
where $\bar{\rho}_{\phi}$ is the background energy density of a homogeneous
scalar field $\bar{\rho}_{\phi}=K(\phi)+V(\phi)$ and $K(\phi)$ its
kinetic energy. In linear theory, the homogeneous term would be the
only term entering into $\Omega_{\phi}$; on the contrary, within
the N-body simulation, the field is non-homogeneous and the combined
energy density of the coupled neutrino-cosmon fluid receives also
a contribution from the perturbations $\delta\rho_{\phi}$ of the
non-homogeneous cosmon field, given by eq.\prettyref{eq:perturbed-cosmon}.
The important quantity determining the evolution of a dynamical dark
energy is not the energy density of the cosmon alone, but the energy
density of the combined cosmon-neutrino fluid, given by 
\begin{equation}
\Omega_{\phi+\nu}=\frac{8\pi G}{3H^{2}}\left(\bar{\rho}_{\phi}+\bar{\rho}_{\nu}\right)\,\,.\label{eq:Omega-cosmon+nu}
\end{equation}
The average energy density of the neutrinos is not individually conserved
and its evolution is given by the continuity equation with a coupling
term on the r.h.s. \cite{ayaita_structure_2012,baldi_hydrodynamical_2010}

\begin{equation}
\rho_{\nu}+2\mathcal{H}\rho_{\nu}=-\beta(\phi)\phi'\rho_{\nu}\,\,,
\end{equation}
where $\phi'$ is the time derivative of the field with respect to
conformal time $\tau$. The corresponding equation of state of the
coupled fluid can then be defined as the sum of the pressure components
divided by the sum of the density components

\begin{equation}
w_{\nu+\phi}=\frac{\bar{p}_{\phi}+\bar{p}_{\nu}}{\bar{\rho}_{\phi}+\bar{\rho}_{\nu}}\,\,.\label{eq:Wqnu}
\end{equation}
In the literature \cite{brookfield_cosmology_2007,das_super-acceleration_2006,perrotta_extended_1999,perrotta_dark_2002},
there are several definitions of the effective equation of state or
the observed equation of state in the case in which the scalar field
is coupled to other particles. We argue that \prettyref{eq:Wqnu}
is actually the equation of state one would observe from the evolution
of the Hubble function (i.e. with Supernovae and standard candle methods
of redshift distance measurements). In appendix \ref{sec:The-effective-W}
we comment further on this and show a comparison between the ``observed''
and theoretical equation of state of dark energy in fig.\ref{fig:Wnuphi-Wde-model2}.

\begin{figure}
\centering{}\includegraphics[width=0.75\textwidth]{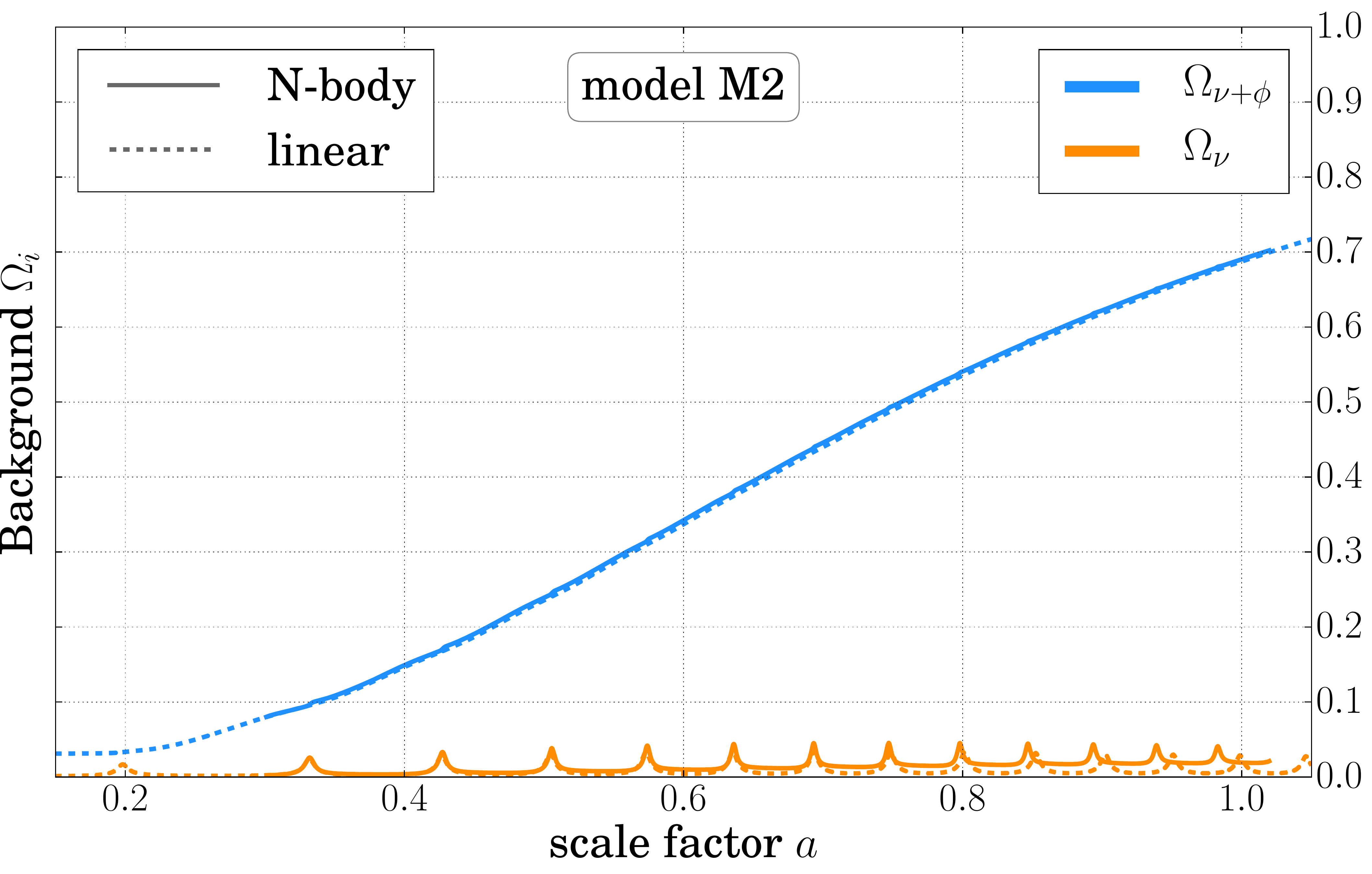}\protect\caption{\label{fig:bckg-mod2-1}Evolution of $\Omega_{\phi+\nu}$ (blue lines)
and $\Omega_{\nu}$ (orange lines) for Model M2, compared between
the linear output from nuCAMB (dashed lines) and the non-linear calculation
of the N-body simulation (solid lines). The total cosmon-neutrino
fluid has the same background evolution in the simulation as in the
linear calculation. The neutrino energy density is somewhat larger
in the simulation and shows a phase-shift in its oscillations, as
discussed in the text.}
\end{figure}

In fig.\ref{fig:bckg-mod2-1} we plot for model M2 the background
evolution of the neutrino energy density $\Omega_{\nu}$ (orange lines)
and the combined cosmon+neutrino fluid energy density $\Omega_{\nu+\phi}$
(blue lines) as defined in eq.\prettyref{eq:Omega-cosmon+nu}. The
dashed lines correspond to the linear computation in nuCAMB, while
the solid lines correspond to the results of the N-body simulation.
One can see that the effect of non-linearities and backreaction is
quite small and it is mostly just visible as a phase shift in the
oscillations of $\Omega_{\nu}$, which is due to the dynamics of the
oscillating lumps, that alter the field-dependent mass of the neutrinos
as a function of time and space. The same trend is observed in model
M1 (not shown here). This behavior tells us that for small neutrino
masses, the effects of backreaction on the background evolution are
practically negligible and a linear computation is enough to analyze
those models further, with a considerable simplification with respect
to a joint linear and non-linear analysis done in \cite{pettorino_neutrino_2010}.

\begin{figure}
\centering{}\includegraphics[width=0.75\textwidth]{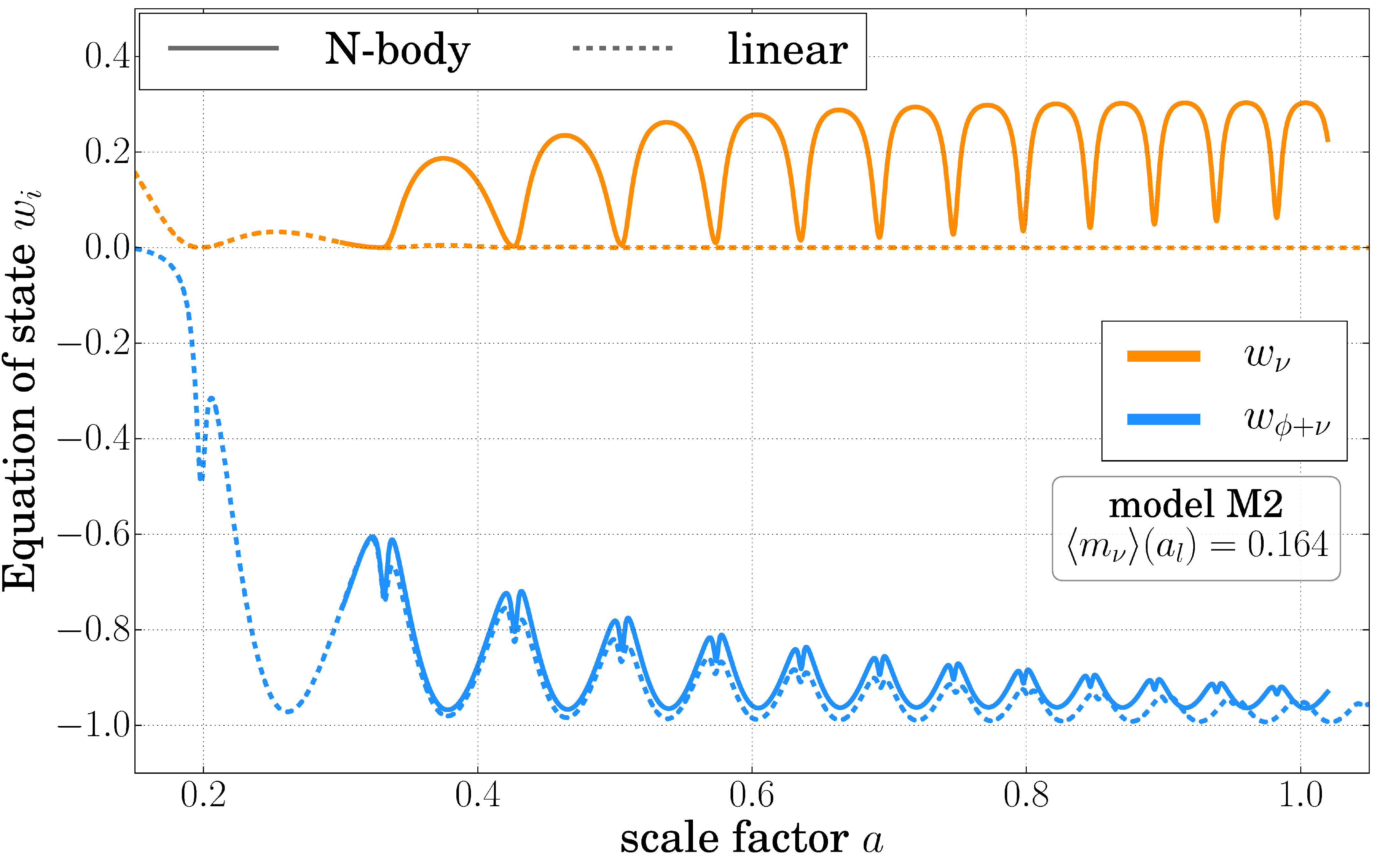}\protect\caption{\textbf{\label{fig:backg-wnu-wqnu-mod2}} Equation of state of the
combined neutrino-cosmon fluid $w_{\phi+\nu}$ (blue) and equation
of state of neutrinos $w_{\nu}$ (orange). We compare the linear output
(dashed lines) to the non-linear one obtained from the N-body simulation
(solid lines) for model M2. This model has a time averaged RMS mass
$\langle m_{\nu}\rangle(a_{l})=0.164$, where $a_{l}=0.9$ in the
label denotes the center of the time interval $a=[0.8-1.0]$ used
to take the average. For $w_{\nu}$, the linear output does not capture
the oscillating equation of state of neutrinos due to the formation
of structures, while for $w_{\phi+\nu}$, both codes agree relatively
well. At late times the equation of state predicted by the simulation
has a somewhat higher value and is phase-shifted due to the heating
of the neutrino fluid.}
\end{figure}

We show in fig.\ref{fig:backg-wnu-wqnu-mod2} the neutrino equation
of state $w_{\nu}$ (orange lines) as well as the combined cosmon-neutrino
fluid equation of state $w_{\nu+\phi}$ as defined in equation \ref{eq:Wqnu}
(blue lines), both for the case of the linear computation with nuCAMB
(dashed lines) and the non-linear computation (solid lines). In the
linear analysis, the neutrinos are treated initially as relativistic
particles: as the mass increases, they become more and more non-relativistic,
reaching a $w_{\nu}$ of exactly zero at late times. On the contrary,
the N-body simulation is able to follow the oscillations in the equation
of state of neutrinos, which are caused by the fact that neutrinos
get accelerated to relativistic velocities when they fall into deep
gravitational and cosmon potentials. Once they are in these lumps,
and they have acquired high speeds, their pressure increases and they
tend to escape again from these lumps, causing the oscillating neutrino
structures. When they are far away from the cosmon potentials, their
velocities decrease and they become non-relativistic again. The fifth
force acting among neutrinos attracts them again to the cosmon potential
wells and the whole cycle repeats itself.

For the combined equation of state $w_{\nu+\phi}$, we find that the
simulation predicts a slightly higher value than the linear one; this
can be explained by studying how the neutrino fluid is heated due
to the strong oscillations of the lumps. By falling repeatedly in
the cosmon potential wells and increasing their kinetic energy, the
neutrinos temperature increases and therefore neutrinos do not manage
to become again completely non-relativistic. This can be seen in the
dashed orange lines of plot \ref{fig:backg-wnu-wqnu-mod2}, where
the curve of $w_{\nu}$ does not touch the zero axis after $a\approx0.5$.
We will see in section \ref{sec:heating-of-the-neutrino} that neutrinos
depart from their initial Fermi-Dirac distributions and reach temperatures
which are high compared to the photon background.

\begin{figure}
\begin{centering}
\includegraphics[width=0.75\textwidth]{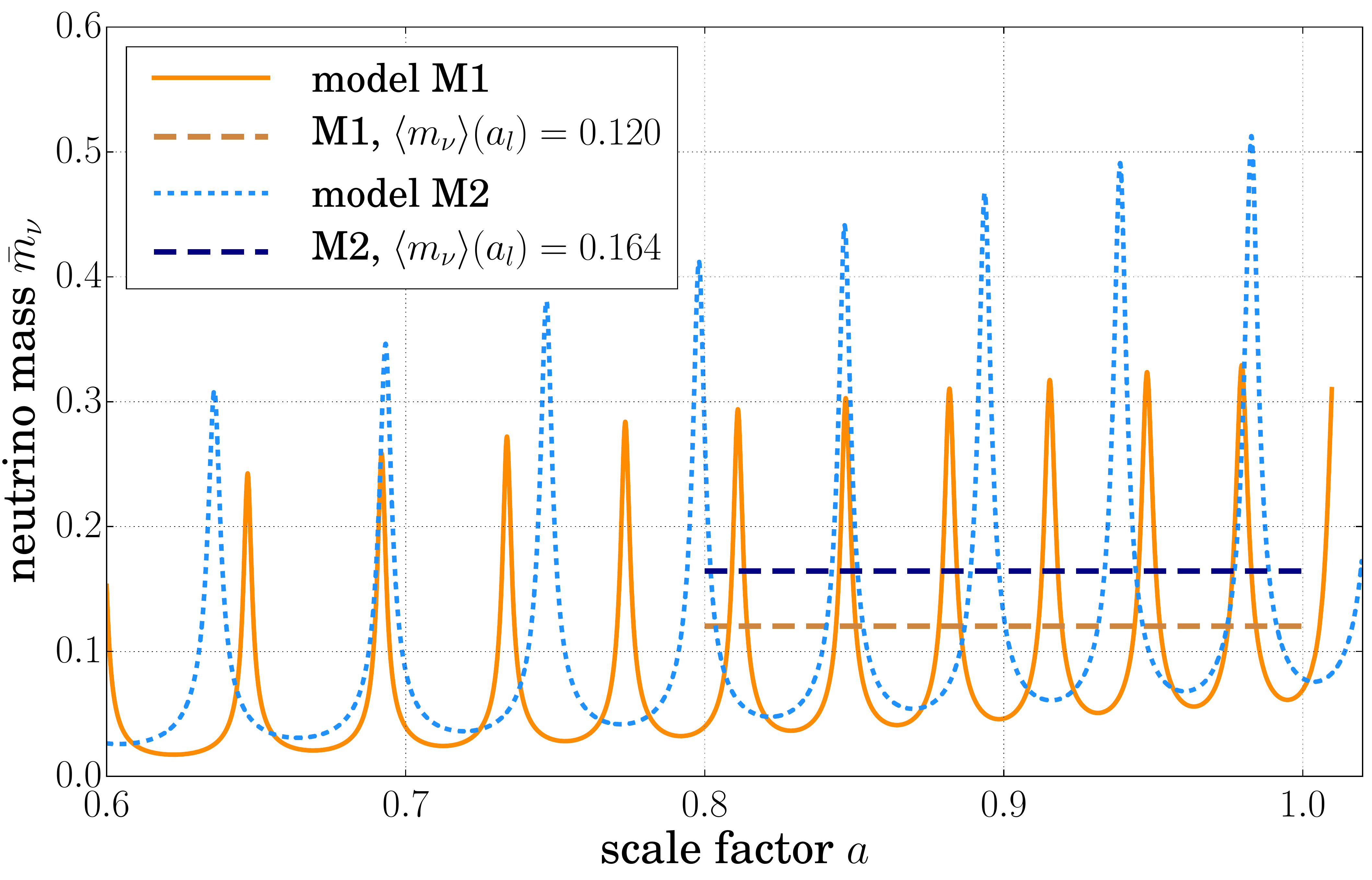} 
\par\end{centering}

\protect\protect\protect\protect\protect\protect\caption{\textbf{\label{fig:NuMass-mod2}}Neutrino mass $\overline{m}_{\nu}$
(average over simulation volume) in model M2 (firebrick red line)
and model M1 (dodger blue line), as a function of the scale factor
$a$, for the N-body simulation. The horizontal lines, show the time
averaged RMS value at a late time $a_{l}=0.9$, denoting the center
of the time interval of $[0.8:1.0]$ considered for taking the average.
For model M2, the time averaged neutrino mass is $\langle m_{\nu}\rangle(a_{l})=0.164$
(blue dashed lines), while for model M1 it is somewhat smaller $\langle m_{\nu}\rangle(a_{l})=0.120$
(red dashed lines). One can observe that the oscillation frequency
is higher for the smaller $\langle m_{\nu}\rangle$ mass and the peaks
are higher for the larger $\langle m_{\nu}\rangle$. The present neutrino
masses of the two models calculated in linear theory differ by an
order of magnitude, on the contrary the time averaged masses are very
close to each other. }
\end{figure}

In fig.\ref{fig:NuMass-mod2}, we show the evolution of the spatial
average of the neutrino mass in the N-body simulation as a function
of the scale factor $a$. One can see that the value of $\bar{m}_{v}$
varies along an order of magnitude, from approx.$10^{-2}$ to $10^{-1}$,
throughout a cosmological time interval. Due to a phase shift in the
oscillation pattern, which sets in at around $a\approx0.8$, the present
day value of the average neutrino mass can be quite different to the
one estimated with the linear analysis (and this change depends on
the precise parameters of the model), so that the best estimate for
the average cosmological neutrino mass today, is a time average of
$\bar{m}_{v}(\phi)$ at late times, between $a=0.8-1.0$. We can see
that the big discrepancies between the masses of model M1 and M2 calculated
in linear theory (e.g. \prettyref{tab:simus-params}) are washed away
when non-linearities and backreaction effects are taken into account
i.e. for small neutrino masses. Even if the present neutrino masses
for model M1 and M2 differ by an order of magnitude in linear theory,
we find a very similar time averaged value between the two models
in the N-body simulation: respectively $\langle m_{\nu}\rangle[0.8:1.0]=0.120$
and $\langle m_{\nu}\rangle[0.8:1.0]=0.164$. The oscillation pattern
of the neutrino mass for the more massive model (M2) contains higher
peaks and has a smaller frequency than compared to the oscillations
in the less massive model M1. For other models, this comparison can
be seen in table \prettyref{tab:simus-params}.

\section{\label{sec:heating-of-the-neutrino}heating of the neutrino fluid}

The repeated acceleration of neutrinos to relativistic velocities
during the periods of lump formation and dissolution lead to an effective
heating of the neutrino fluid. While we do not expect a thermal equilibrium
distribution of neutrino momenta and energies it is interesting to
investigate how close the distribution is to the Fermi-Dirac distribution
of a free gas of massive neutrinos. This distribution depends on only
two parameters, the neutrino mass and the temperature. At a given
time we associate the neutrino mass to the space averaged neutrino
mass. The temperature can be associated to the mean value of the momentum.

The energy of a relativistic particle is given by

\begin{equation}
E(p,m)=\sqrt{p^{2}+m^{2}}\,\,,\label{eq:rela-energy}
\end{equation}
while its kinetic energy $E_{k}=E(p,m)-m$. Equivalently, the kinetic
energy is defined by

\begin{equation}
E_{k}=\int\vec{v}\:\cdot\mbox{d}\vec{p}\,\,,
\end{equation}
which yields $E_{k}=m(\gamma-1)$ and reduces in the limit of very
small velocities ($v\lll c$) to the usual $E_{k}=mv^{2}/2$. From
there the Fermi-Dirac distribution as a function of momentum $p=|\vec{p}|$
can be obtained in the standard way. It depends on $m$ and $T$.
For $m\ll T$ it can be approximated by the relativistic distribution
while for $m\gg T$ we recover the Maxwell-Boltzmann distribution.
The distribution of particle momenta is then given by

\begin{equation}
\mathcal{P}(p)\mbox{d}p=\frac{4\pi p^{2}}{(1+e^{(E(p,m)-\mu)/T})}\mbox{d}p\label{eq:momentum-dist}
\end{equation}
where the factor $4\pi$ comes form the angular integration of the
three-dimensional momentum. In the ultra-relativistic limit we can
analytically integrate the momentum $p$ over its distribution eq.\prettyref{eq:momentum-dist}
and invert $\bar{p}(\overline{T})$ to yield the mean temperature
as a function of the mean momentum

\begin{equation}
\bar{T}=\frac{180\zeta(3)}{7\pi^{4}}\bar{p}
\end{equation}
We neglect the chemical potential in \prettyref{eq:momentum-dist},
because the exponential term in the denominator is 2 or 3 orders of
magnitude larger than unity. Since in our case, the average momentum
and mass of the neutrinos are of the same order, we cannot use either
a non-relativistic or an ultra-relativistic limit. We need to consider
both the mass and the momentum in the relativistic energy equation
\prettyref{eq:rela-energy}. Therefore for each model and each time,
we numerically find $\bar{T}$ as a function of the mean momentum
$\overline{p}$.

We extract for $a=1$ the temperatures 
\begin{equation}
\overline{T}=0.077\mbox{eV}\,\,\mbox{(M1, }\bar{m}_{\nu}=0.2404\mbox{)},\,\,\overline{T}=0.065\mbox{eV}\,\,\mbox{(M2, }\bar{m}_{\nu}=0.2327\mbox{)}\,\,.
\end{equation}
They are higher by a factor 327 (M1) or 276 (M2) as compared to the
CMB photon temperature $2.35\times10^{-4}$eV. This demonstrates the
unconventional heating of the neutrino fluid due to the formation
and dissolution of lumps. The high temperatures are connected with
the almost relativistic equation of state of the neutrinos seen in
fig. \ref{fig:backg-wnu-wqnu-mod2}. Overall, the observed momentum
distributions come rather close to the thermal equilibrium distribution.
This also holds for the distribution of kinetic energies. With the
bulk quantities as momenta and kinetic energies roughly distributed
thermally this is an example of prethermalization \cite{berges_prethermalization_2004}.

\begin{figure}
\begin{centering}
\includegraphics[width=0.49\textwidth]{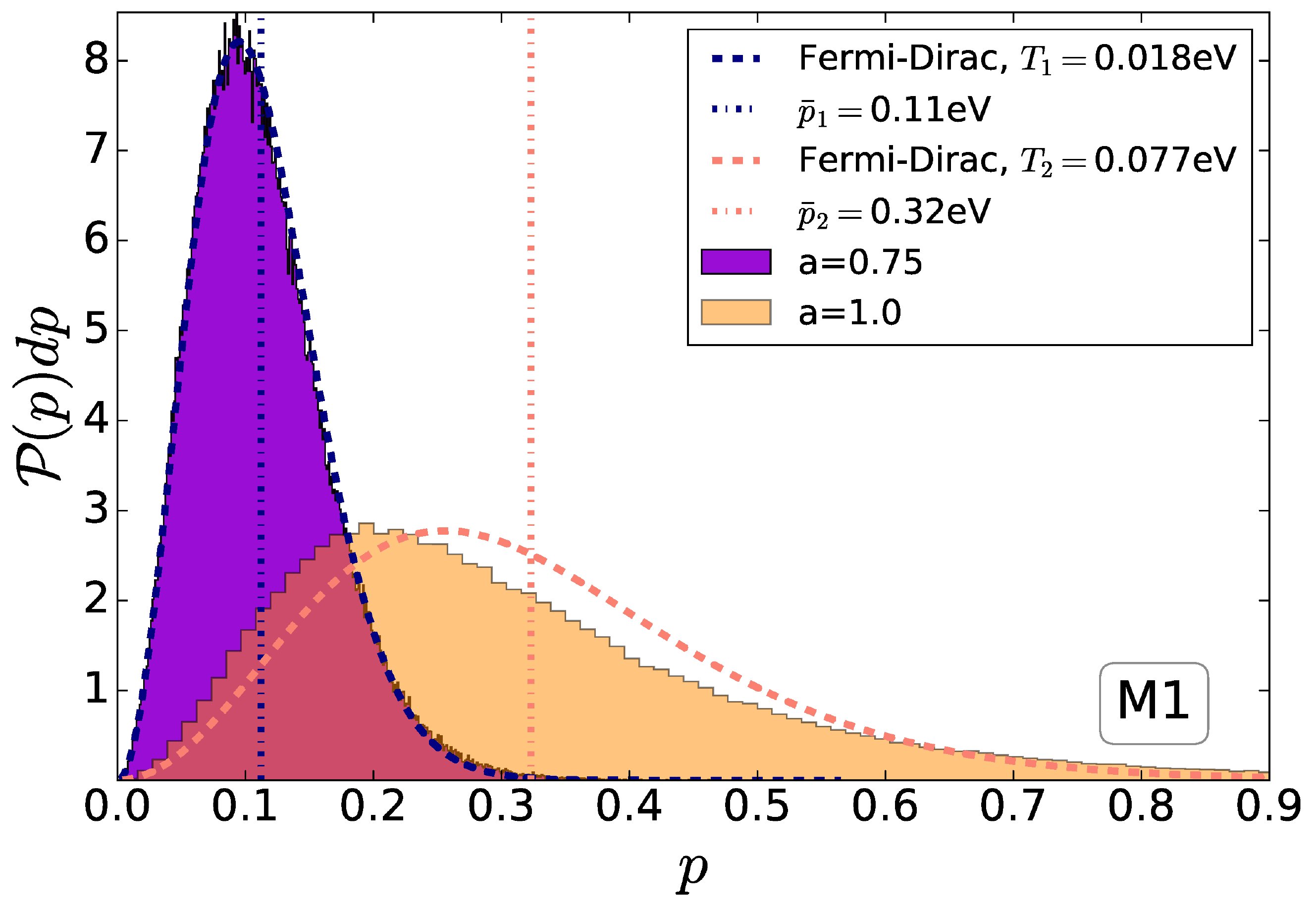}\includegraphics[width=0.49\textwidth]{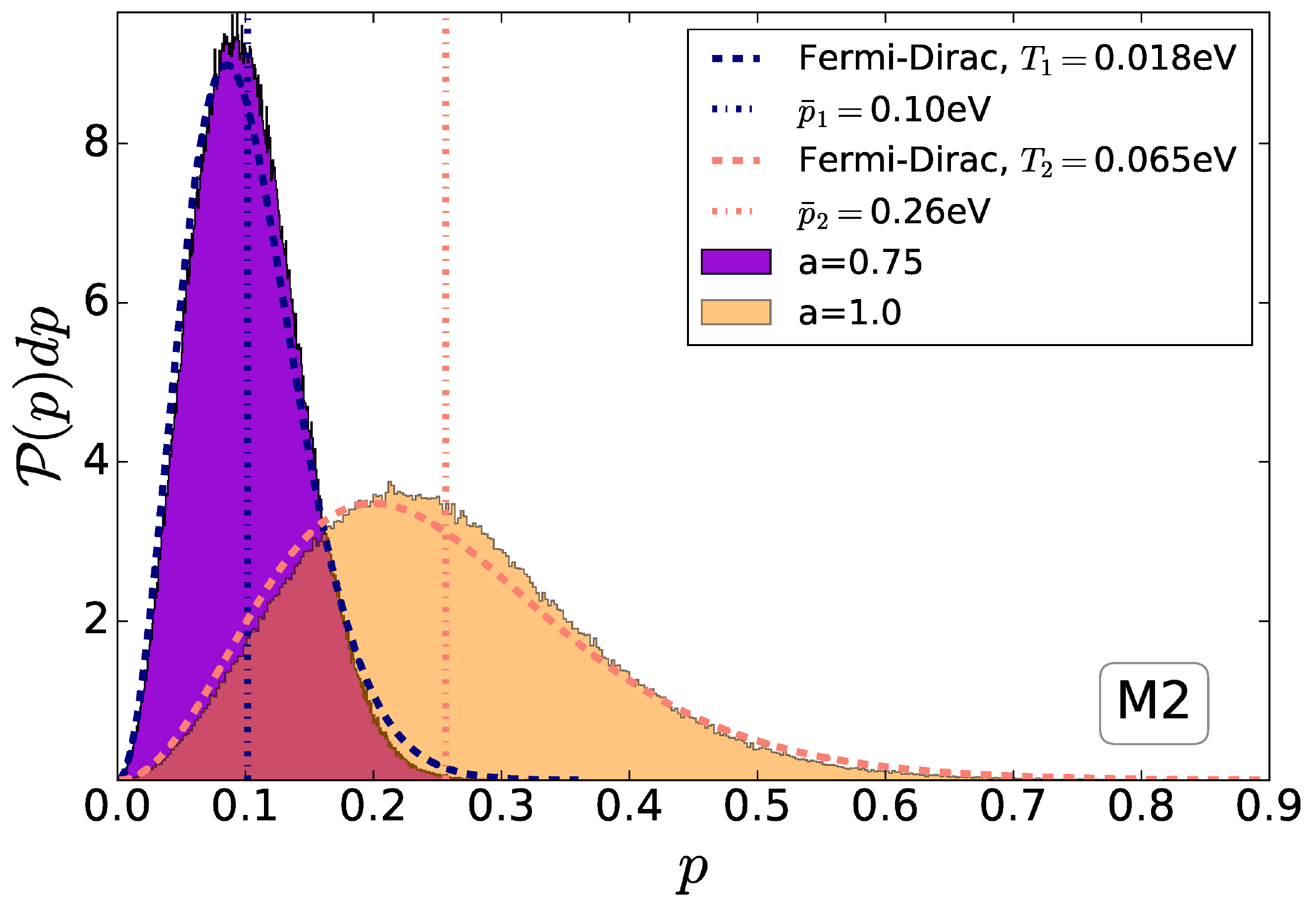} 
\par\end{centering}

\protect\protect\caption{\label{fig:histogram-FermiD-1} Distribution of the momenta of the
neutrino particles in the simulation, for two different times, $a=0.75$
(purple shade) and $a=1.0$ (orange shade), compared against a Fermi-Dirac
distribution with a temperature given by the mean of the distribution
(dashed lines). \textbf{Left: }For model M1 the Fermi-Dirac fits very
well for temperatures of $\bar{T}=0.018$eV and $\bar{T}=0.077$eV
for each scale factor respectively. \textbf{Right: }For model M2 the
fit is also good, the corresponding temperatures being $\bar{T}=0.018$eV
and $\bar{T}=0.065$eV. The CMB photon temperature is $2.35\times10^{-4}$eV,
this means that the non-linear cosmon-neutrino interactions heat the
neutrino background by more than a factor 100.}
\end{figure}

In fig. (\ref{fig:histogram-FermiD-1}) we fit the distribution of
momenta of the neutrino particles on the grid (shown with an histogram)
with a Fermi Dirac distribution. The actual distribution of momenta
fits the thermal equilibrium distribution very well. At later times
(orange shade), the fit is slightly less good: neutrinos might be
accelerating towards or away from lumps giving them an extra kick
that shifts the peak of the distribution of momenta.

When comparing the equation of state of neutrinos obtained from the
N-body simulation to a neutrino equation of state $w_{\nu}=p_{\nu}/\rho_{\nu}$,
using our Fermi-Dirac fit to the particle distribution and eqns.\prettyref{eq:pnu-FD}
and \prettyref{eq:rhonu-FD}, we get a very good agreement, taking
into account that for the Fermi-Dirac fit, we are neglecting the spatial
variation of the neutrino mass $m_{\nu}(\phi)$. For model M2 at the
scale factor $a=0.75$ we obtain from the N-body simulation a neutrino
equation of state of $w_{\nu}=0.081$ while using the Fermi-Dirac
fit to the distribution of particles with a mean temperature of $\bar{T}=0.018$eV
and an average neutrino mass $\bar{m}_{\nu}=0.1835$, the proper calculation
yields $w_{\nu}=0.086$. For a later time, at $a=1.0$ the N-body
simulation gives us a value of $w_{\nu}=0.207$ while the Fermi-Dirac
fit with a mean temperature of $\bar{T}=0.065$eV and an average neutrino
mass $\bar{m}_{\nu}=0.2327$ amounts to a neutrino equation of state
of $w_{\nu}=0.182$.

To visualize the evolution of $w_{\nu}$, we can observe from fig.\ref{fig:backg-wnu-wqnu-mod2}
that neutrinos in the N-body simulation start as non-relativistic
particles and oscillate between being almost relativistic and completely
non-relativistic in the interval $a=[0.3,0.6]$. However, at later
times $a\apprge0.7$ the neutrino equation of state still oscillates
but never reaches a value of zero again. This is in agreement with
our description of the heating of the neutrino fluid. Since the mean
temperature of the neutrino fluid is increasing with time and therefore
its mean kinetic energy and pressure, the minimum of the oscillations
of the neutrino equation of state increases also in time and departs
from zero, once neutrinos are heated to very high temperatures due
to the collapsing and dissolving of the neutrino-cosmon lumps.

\begin{figure}
\centering{}\includegraphics[width=0.7\textwidth]{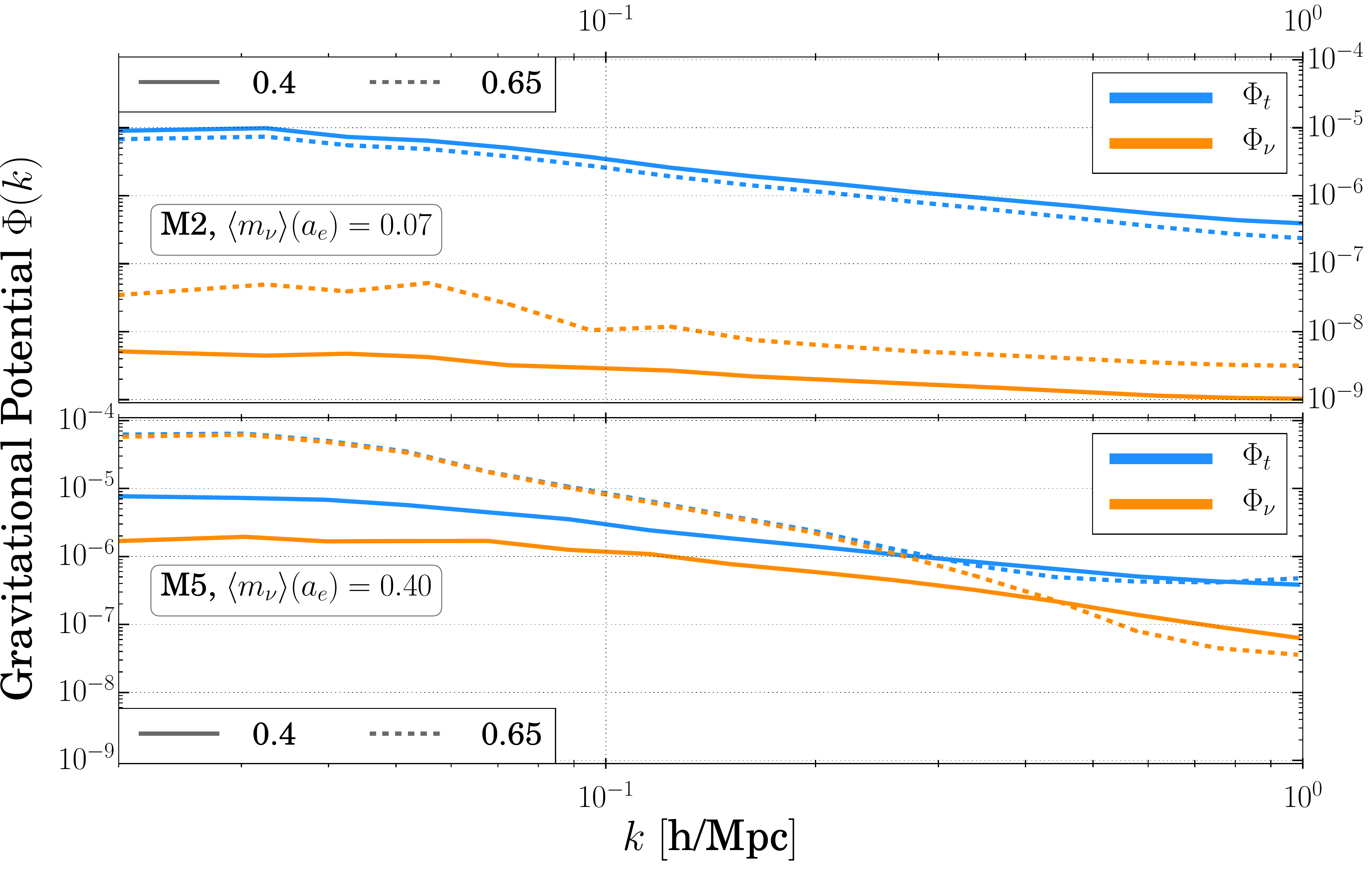}\protect\caption{\label{fig:gravpot-in-scale} Power spectra of the total gravitational
potential $\Phi_{t}$ (blue lines) and of the neutrino contribution
$\Phi_{\nu}$ (orange lines) for model M2 and model M5 at the scale
factors $a=0.40$ (solid lines) and $a=0.65$ (dashed lines), as a
function of scale. Model M2 has an RMS time averaged neutrino mass
$\langle m_{\nu}\rangle(a_{e})=0.07$, where $a_{e}=0.5$ stands for
the central time in the interval $a=[0.4-0.6]$ used to take the average.
Model M5 has in the same interval a higher RMS mass of $\langle m_{\nu}\rangle(a_{e})=0.40$.
In the first model, $\Phi_{t}$ at large scales is of the order of
$10^{-5}$, while $\Phi_{\nu}$ is 3 to 4 orders of magnitude smaller
at both cosmological times. For model M5, in which neutrino lumps
are stable and growing, one sees that at large scales, the total $\Phi_{t}$
starts with a value of $10^{-5}$ at $a=0.4$, but reaches $10^{-4}$
at later times. At $a=0.65$ the neutrino contribution is dominant
and neutrino structures have migrated from small scales to large scales,
as can be seen from the dip in $\Phi_{\nu}$ at modes between $k=0.2-1.0$
h/Mpc.}
\end{figure}

\section{Gravitational Potentials of Neutrino Lumps}

\begin{figure}
\centering{}\includegraphics[width=0.7\textwidth]{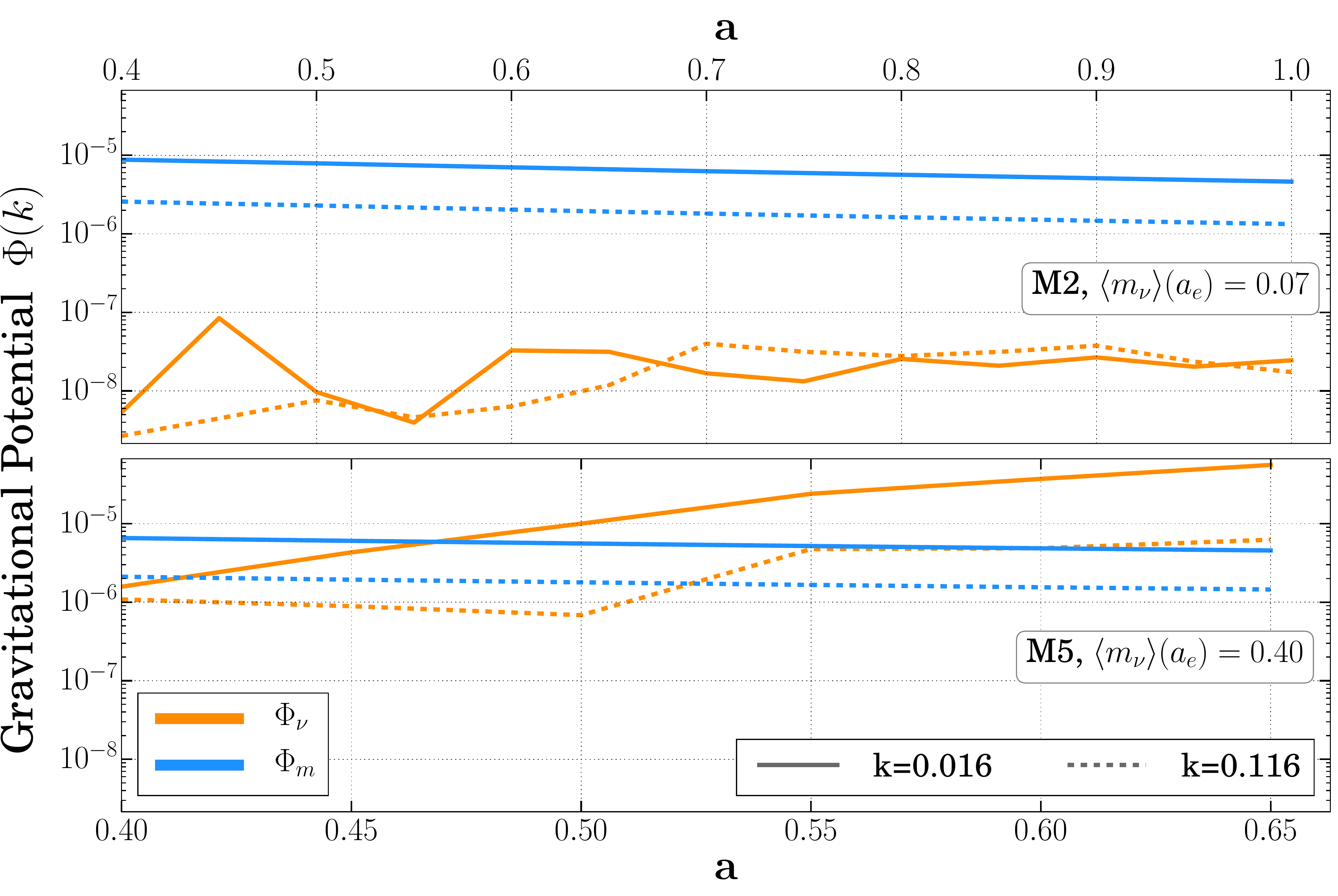}\protect\caption{\label{fig:gravpot-intime1} Power spectra of the matter gravitational
potential $\Phi_{m}$ (blue lines) and of the neutrino contribution
$\Phi_{\nu}$ (orange lines) at two different scales, $k=0.016$ (solid
lines) and $k=0.116$ (dashed lines) as a function of scale factor
$a$ with two different scales on the top and bottom axis. For model
M2 the matter gravitational potential $\Phi_{m}$ is at most $10^{-5}$
at all times, while the neutrino contribution is 2-3 orders of magnitude
smaller and displays time oscillations. In clear contrast, the neutrino
contribution from model M5 for very large scales, reaches and dominates
over the matter contribution for $a\apprge0.5$ and pushes the total
$\Phi$ to high values that would be ruled out by observations, see
also fig. \ref{fig:gravpot-in-scale}. The RMS neutrino mass has been
taken in the same interval range as for fig.\ref{fig:gravpot-in-scale},
where $a_{e}=0.5$.}
\end{figure}

The gravitational potential $\Phi$ is a good measure of the physics
going on in structure formation. We know from observational constraints,
that $\Phi$ is of the order of $10^{-5}$ on cosmological scales
\cite{pettorino_neutrino_2010,brouzakis_nonlinear_2011}. In $\lcdm$,
the gravitational potential is sourced mainly by dark matter perturbations.
In figures \ref{fig:gravpot-in-scale} and \ref{fig:gravpot-intime1}
we show that for models with small neutrino masses, the neutrino contribution
to $\Phi$ remains several orders of magnitude smaller than the CDM
contribution, at all scales and at all times. Moreover, one can observe
an oscillation in time of the neutrino gravitational potential. For
models with large neutrino masses, the neutrino contribution grows
monotonically with time. At large scales $k\lesssim0.3$h/Mpc and
at late times the neutrino lump induced potential dominates over the
cold dark matter gravitational potential. This renders the total potential
$\Phi_{tot}$ too big to be compatible with present cosmological constraints.

We show the scale dependence of the total gravitational potential
and the neutrino induced gravitational potential $\Phi_{\nu}$ at
two different cosmic time scales $a=0.4$ and $a=0.65$ in fig.\ref{fig:gravpot-in-scale}.
While for $a=0.4$ (solid lines) the neutrino contribution is still
subdominant for both models M2 and M5, this changes at $a=0.65$ for
model M5 (bottom panel). For model M2 the total gravitational potential
decreases in time, as it is expected due to the effect of dark energy,
while $\Phi_{\nu}$ increases especially at large scales. For model
M5, since the neutrino contribution dominates at $a=0.65$, the total
gravitational potential is raised to values of $10^{-4}$ at large
scales, while at small scales ($k\apprge0.4$) the neutrino contribution
is still subdominant.

\begin{figure}
\centering{}\includegraphics[width=0.65\textwidth]{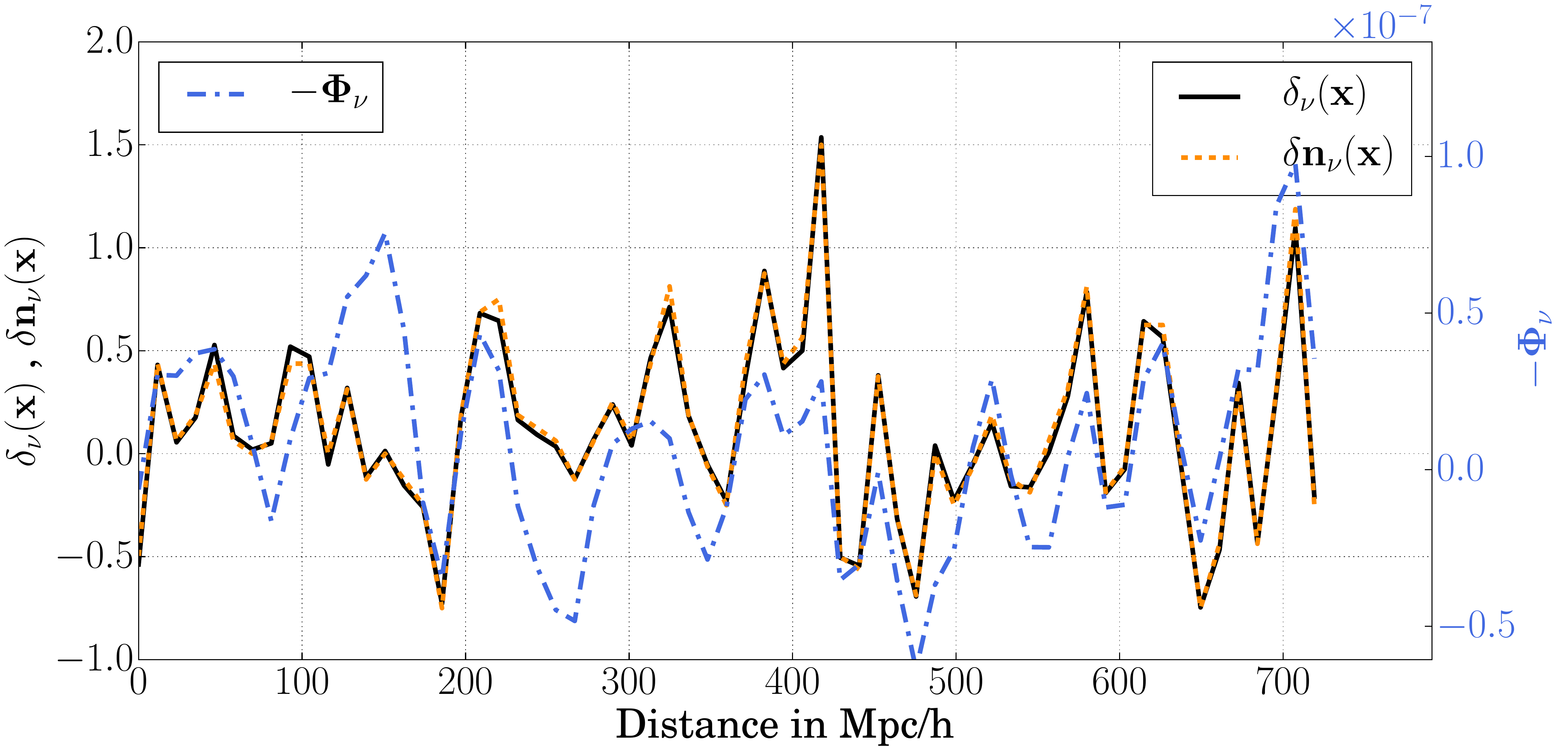}\includegraphics[width=0.35\textwidth]{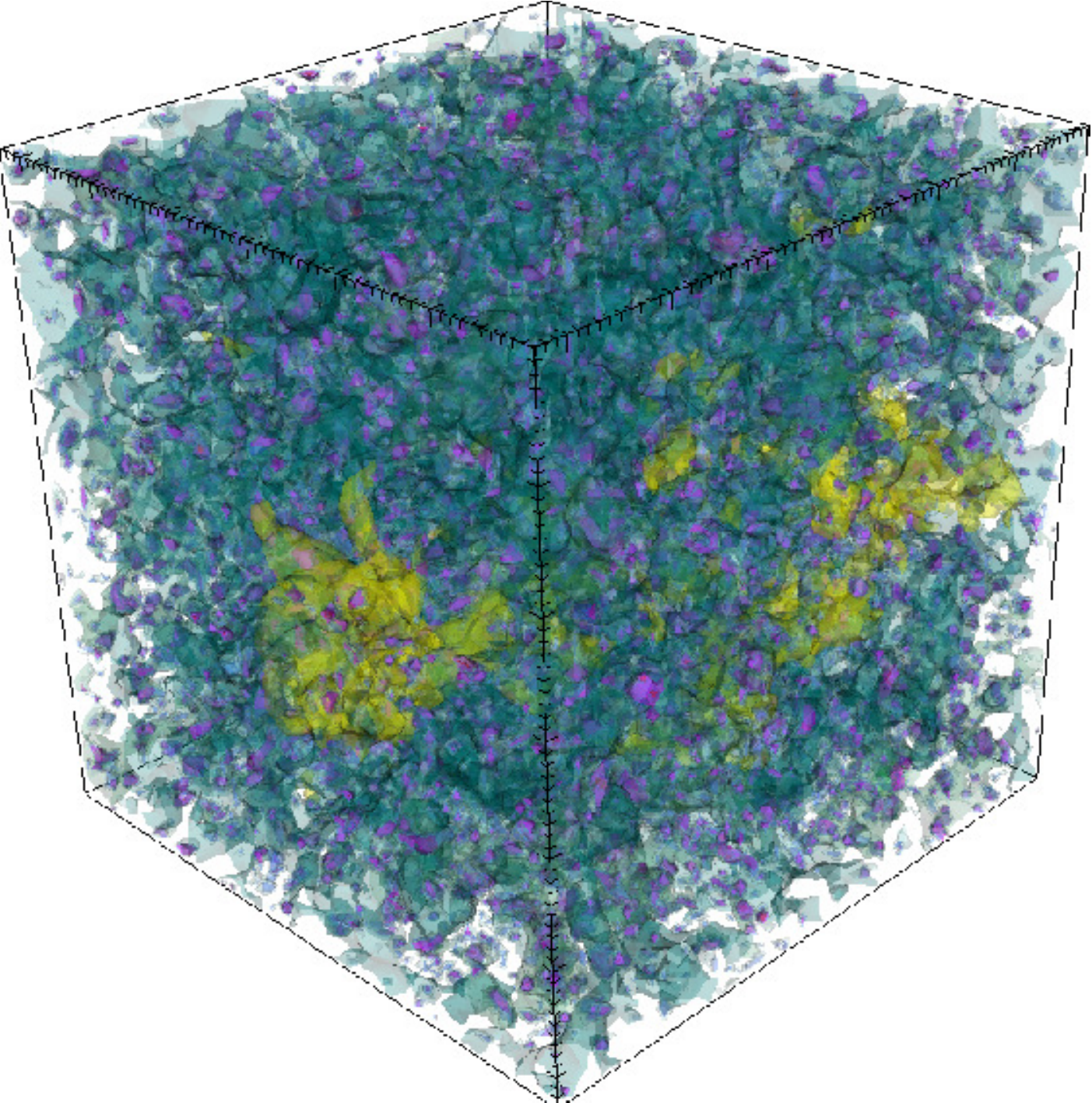}\protect
\caption{\label{fig:lineout-gravpotnu} 
\textbf{Left: }Line plot through the
main diagonal of the simulation box, for model M2 at $a=0.75$. The
negative of the neutrino contribution to the gravitational potential
$\Phi_{\nu}$ (blue dot-dashed lines) oscillates between $\pm1.0\times10^{-7}$.
This is correlated to the neutrino density contrast (black lines)
and the number density contrast $\delta n_{\nu}=n_{\nu}(\vec{x})/\bar{n}_{\nu}-1$
(orange dashed lines) reaching values of up to 1.5. \textbf{Right:
}Snapshot of the same simulation, showing a equipotential contour
of the gravitational potential, for $\Phi_{\nu}=+1.0\times10^{-7}$
in yellow, and the small but dense neutrino lumps in blue, purple
and red, corresponding to density contrasts $\delta n_{\nu}$ of 1.5,
2.0 and 4.0 respectively.}
\end{figure}

In fig.\ref{fig:gravpot-intime1} we show the power spectra of the
total gravitational potential $\Phi$ (blue lines) and of the neutrino
contribution to it (orange lines) at two different scales, $k=0.016$
(solid lines) and $k=0.116$ (dashed lines) as a function of the scale
factor $a$. For model M2, corresponding to an average early time
neutrino mass of 0.07 eV, the total gravitational potential $\Phi(k)$
is $10^{-5}$ at all times, while the neutrino contribution is 2-3
orders of magnitude smaller and shows time oscillations. In clear
contrast, the neutrino contribution from model M5 (corresponding to
an average early time neutrino mass of 0.40 eV)) for very large scales
reaches and dominates over the matter contribution (at $a\apprge0.5$)
and pushes the total $\Phi$ to high values that would be ruled out
by observations. This is due to the fact that neutrino lumps do not
dissolve, but rather grow with continuously growing concentration
and higher gravitational potential.

There is also anticorrelation between the neutrino structures and
the neutrino induced gravitational potential, as expected from the
fact that neutrinos will tend to fall into gravitational potential
wells. In the left panel of fig.\ref{fig:lineout-gravpotnu}, we plot
the values of the neutrino number density contrast $\delta n_{\nu}=n_{\nu}(\vec{x})/\bar{n}_{\nu}-1$
and the negative neutrino induced gravitational potential $\Phi_{\nu}$,
along a diagonal line through the simulation box. The correlation
of peaks and troughs (corresponding to an anticorrelation of $\delta n_{\nu}$
and $\Phi_{\nu}$) is very clear and it is valid for even small substructures
of the order of a few Mpc. By plotting the neutrino density contrast
$\delta_{\nu}$, we also show that at this time $a=0.75$, the neutrino
number density and the energy density are proportional, meaning that
neither local mass variations or relativistic speeds are having any
effect in the neutrino total energy. In the right panel of fig.\ref{fig:lineout-gravpotnu}
we visualize the neutrino induced gravitational potential as a yellow
region marking the equipotential surface $\Phi_{\nu}=+1.0\times10^{-7}$
and the neutrino number overdensity structures colored blue, purple
and red, corresponding to density contrasts $\delta n_{\nu}$ of 1.5,
2.0 and 4.0 respectively. For this model (M2) and at this specific
time, the neutrino structures are spread almost homogeneously throughout
the simulated volume.

\FloatBarrier

\section{Conclusions}

We have investigated the dynamics of neutrino lumps in growing neutrino
quintessence and how it depends on the mass of neutrinos. As a main
result of this paper we found a characteristic divide in the qualitative
behavior between small and large neutrino mass.

For light neutrino masses the combined effects of oscillations in
the neutrino masses and the cosmon-neutrino coupling lead to rapid
formation and dissociation of the neutrino lumps. The concentration
in the neutrino structures never grows to very large overdensities.
As a consequence, backreaction effects remain small. The effects of
lump formation and dissociation lead to an effective heating of the
neutrino fluid to temperatures much higher than the photon temperature.
Due to this heating, the neutrino equation of state becomes again
close to the one for relativistic particles. For a small present average
neutrino mass $m_{\nu}=0.06$eV it has been found earlier \cite{pettorino_neutrino_2010}
that the cosmology of growing neutrino quintessence resembles very
closely a cosmological constant, making differences to the $\lcdm$
model difficult to detect. We extend this qualitative feature to a
whole range of light neutrino masses.

For large neutrino masses, one finds a qualitatively different behavior.
Big neutrino lumps form, due to the strong cosmon-mediated fifth-force
between neutrinos. These lumps are stable and keep growing in concentration
and density. The strong clumping of the cosmic neutrino background
induces large backreaction effects on the overall cosmic evolution.
As a result, the combined cosmon-neutrino fluid does not act effectively
as a cosmological constant anymore and compatibility with observations
is difficult to achieve. This situation is similar to the case of
a constant cosmon-neutrino coupling \cite{fuhrer_backreaction_2015}.

The divide in the characteristic behavior reflects the competition
between heating of the neutrino fluid and lump concentration. We have
not yet established a quantitatively accurate value of the parameter
$\hat{m}_{\nu}$ where the divide is located, since the numerics are
rather time consuming. In principle, this divide will lead to an upper
bound on the present neutrino mass, as seen in terrestrial experiments.
For models in the vicinity of model M2, which seem compatible with
observations so far, spatial average neutrino masses as large as $0.5$eV
can occur at the peak of oscillations, c.f. fig.\ref{fig:NuMass-mod2}.
We note that if we live inside a neutrino lump the neutrino mass will
be reduced as compared to the cosmological value.

We have further computed the strength of the neutrino-induced gravitational
potential. For light masses, this potential is found to be rather
small, rendering a detection of the neutrino lumps difficult. As neutrino
masses increase towards an intermediate mass region, before reaching
the heavy mass range incompatible with observation, the neutrino-induced
gravitational potentials will get stronger. By continuity we expect
that in the intermediate mass region the clumped neutrino background
becomes observable. 
\begin{acknowledgments}
We would like to thank Florian Führer for insightful help and discussions
about the model and its simulation. V.P. and S.C. acknowledge support
from the Heidelberg Graduate School for Fundamental Physics. The authors
acknowledge support by the state of Baden-Württemberg for the computational
capabilities offered through the bwHPC. This research has been supported
by ERC-AdG-290623DFG and through the grant TRR33 ``The Dark Universe''. 
\end{acknowledgments}

 \bibliographystyle{unsrtnat}
\bibliography{gnq-neutrinos-new}

\appendix

\section{\label{sec:Initial-parameters-for-nucamb}Initial parameters for
generating the models in nuCAMB and in the N-Body simulation}

\begin{table}[H]
\centering{}%
\begin{tabular}{|c|c|c|c|c|c|c|}
\hline 
CAMB values  & M1  & M2  & M3  & M4  & M5  & M6\tabularnewline
\hline 
\hline 
input: $\Omega_{\nu}h^{2}$  & 0.048  & 0.048  & 0.075  & 0.018  & 0.038  & 0.098\tabularnewline
\hline 
$\tilde{m}_{\nu}$ amplitude factor  & $8.35\times10^{-5}$  & $2.0\times10^{-4}$  & $6.0\times10^{-4}$  & $9.9\times10^{-3}$  & $8.8\times10^{-3}$  & $8.8\times10^{-3}$\tabularnewline
\hline 
input $\hat{r}_{\nu eV}$ factor  & 1.5045  & 1.5045  & 2.3508  & 0.5642  & 1.1911  & 3.0718\tabularnewline
\hline 
$V_{i}$  & $0.99\times10^{-7}$  & $0.99\times10^{-7}$  & $0.99\times10^{-7}$  & $0.99\times10^{-7}$  & $0.99\times10^{-7}$  & $0.99\times10^{-7}$\tabularnewline
\hline 
\end{tabular}\protect\protect\protect\caption{\label{tab:CAMB-inipars-1} Table of initial parameters for each model
computed with nuCAMB. $\tilde{m}_{\nu}$ is the neutrino mass amplitude
used in the simulations, $\hat{r}_{\nu eV}$ is the neutrino mass
units conversion factor between the simulations and nuCAMB. $V_{i}$
is the initial value of the cosmon potential. }
\end{table}

\begin{table}[H]
\centering{}%
\begin{tabular}{|c|c|c|}
\hline 
N-body parameters  & Values  & Meaning\tabularnewline
\hline 
\hline 
$L$  & $428$  & Box side length in Mpc/h\tabularnewline
\hline 
$N_{g}$  & 64  & Grid size per dimension\tabularnewline
\hline 
$N_{pdm}$  & 262144  & Number of dark matter particles in the simulation\tabularnewline
\hline 
$N_{p\nu}$  & 524288  & Number of neutrino particles in the simulation\tabularnewline
\hline 
$\mbox{ngs}_{\mbox{acc}}$  & $1.0\times10^{-5}$  & Numerical accuracy for the NGS solver\tabularnewline
\hline 
\end{tabular}\protect\protect\protect\caption{\label{tab:Nbody-inipars-1-1} Table of numerical parameters for each
N-body simulation run. Several tests were performed varying these
parameters, but the overall behavior for the purposes of this paper
was the same. We also tested model M2 with grid sizes of 128 and 8
times the number of particles, not noticing any qualitative difference
in the dynamics. Only deviations of the order of 10\% in perturbation
quantities were observed when varying the grid size or the number
of particles.}
\end{table}

\section{The effective and observed equation of state of dark energy\label{sec:The-effective-W}}

The equation of state $w$ of a species is defined in terms of its
pressure and density as 
\begin{equation}
w=\frac{p}{\rho}\,\,\,.
\end{equation}
For the case of coupled species, there have been several definitions
in the literature (c.f. \cite{brookfield_cosmology_2007,das_super-acceleration_2006,perrotta_extended_1999,perrotta_dark_2002}).
This is due to the fact that when there is an exchange of energy and
momentum between a particle and a scalar field, the time evolution
of matter does not correspond anymore to the volume dilution rule
$\rho(a)=\rho_{0}a^{-3}$. Therefore the equation of state of the
dark energy field will not be as simple as the equation of state of
a homogeneous scalar field, namely

\begin{equation}
w_{\phi}=\frac{p_{\phi}}{\rho_{\phi}}=\frac{(1/2a^{2})\dot{\phi}^{2}+V(\phi)}{(1/2a^{2})\dot{\phi}^{2}-V(\phi)}\,\,\,.
\end{equation}
In \cite{das_super-acceleration_2006,brookfield_cosmology_2007} two
definitions of the equation of state (e.o.s) of dark energy were investigated.
The first one, called the effective e.o.s. $w_{eff}$, given by

\begin{equation}
w_{eff}\equiv w_{\phi}+\frac{\beta\dot{\phi}}{3H}\frac{\rho_{\nu}}{\rho_{\phi}}
\end{equation}
and the second one named ``apparent'' e.o.s. $w_{ap}$, defined
by 
\begin{equation}
w_{ap}\equiv\frac{w_{\phi}}{1+x}\,\,\,,
\end{equation}
with 
\begin{equation}
x=-\frac{\rho_{\nu,0}}{a^{3}\rho_{\phi}}\left[\frac{m_{\nu}(\phi)}{m_{\nu}(\phi_{0})}-1\right]\,\,\,,
\end{equation}
where a $0$ subscript denotes quantities at $z=0$. Notice that by
construction, $w_{ap}$ at the present epoch is identical to $w_{\phi}$
and that $w_{ap}$ can be smaller than $-1$. We find that none of
these two definitions of e.o.s. describe the dynamical dark energy
field present in our model.

\begin{figure}
\centering{}\includegraphics[width=0.85\textwidth]{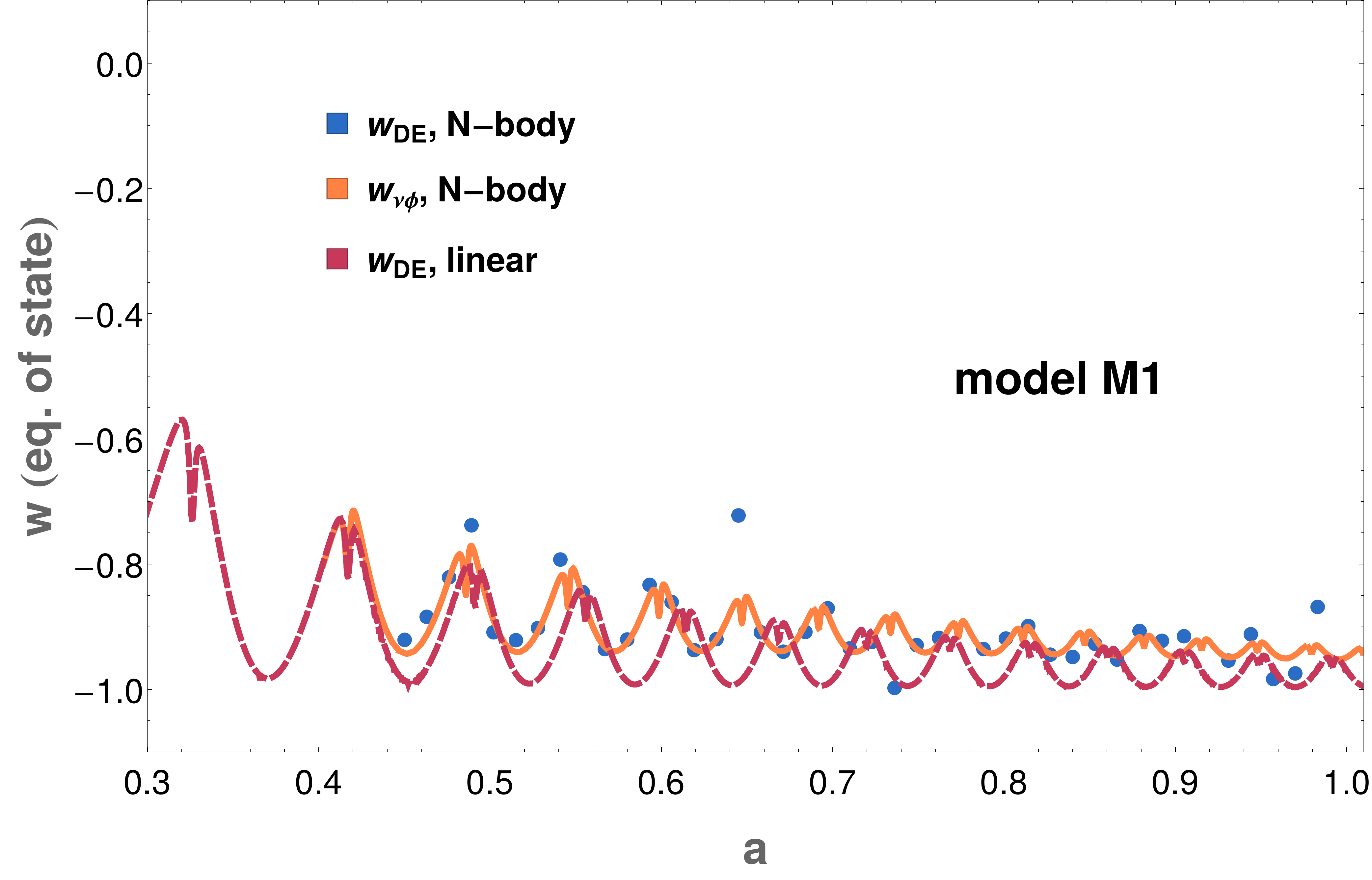}\protect\protect\protect\caption{\label{fig:Wnuphi-Wde-model2} Different $w(z)$ curves in the case
of model M1 computed with the N-body simulation and compared to linear
theory. The red dashed lines represent $w_{DE}$ computed from the
Hubble function which is an output of the linear code nuCAMB. The
solid orange line, representing $w_{\nu+\phi}$ is computed using
the standard pressure and density outputs from the simulation. The
blue dots stand for some set of computed values of $w_{DE}(z)$ obtained
from the Hubble function calculated entirely within the N-body simulation.
Due to numerical noise in the oscillating derivatives of $E(z)$,
there is some scatter in the e.o.s. obtained in this case.}
\end{figure}

Since in our model the neutrino and cosmon field behave together as
a tightly coupled fluid, we are interested in the conserved equation
of state for the combined fluid. This is given by the sum of both
contributions from the pressure, divided by the sum of both densities.
Therefore, we define $w_{\nu+\phi}$ as

\begin{equation}
w_{\nu+\phi}\equiv\frac{\bar{p}_{\phi}+\bar{p}_{\nu}}{\bar{\rho}_{\phi}+\bar{\rho}_{\nu}}\,\,\,.\label{eq:Wqnu-1}
\end{equation}
This definition should agree with what we can extract directly from
observations of the Friedmann equation. The equation of state $w_{DE}(z)$
of a general dark energy component that fulfills the continuity equation
(if it is composed by two coupled fluids, the continuity equation
is fulfilled for the sum of densities and pressures), appears in the
Friedmann equation as

\begin{equation}
E^{2}(z)\equiv\frac{H^{2}(z)}{H_{0}^{2}}=\left[\Omega_{r,0}(1+z)^{4}+\Omega_{m,0}(1+z)^{3}+\Omega_{DE,0}\exp\left\{ \int_{0}^{z}\frac{3(1+w_{DE}(\tilde{z}))}{1+\tilde{z}}\mbox{d}\tilde{z}\right\} +\Omega_{k,0}(1+z)^{2}\right]\,\,\,.
\end{equation}
Moreover, in a flat Universe and with a negligible contribution from
radiation, we can solve for $w_{DE}(z)$ and obtain \cite{amendola_dark_2010}:

\begin{equation}
w_{DE}(z)=\frac{(1+z)(E^{2}(z))'-3E^{2}(z)}{3\left[E^{2}(z)-\Omega_{m,0}(1+z)^{3}\right]}\,\,\,.\label{eq:wde-friednmann}
\end{equation}
Thus, from background expansion observations we can obtain a constrain
on $w_{DE}(z)$, provided we know from large-scale structure or CMB
observations the present value of $\Omega_{m}$. In the fig. \ref{fig:Wnuphi-Wde-model2}
we compare $w_{DE}(z)$ obtained from eq.\prettyref{eq:wde-friednmann}
and $w_{\nu\phi}(z)$ computed both consistently within the N-body
simulation. In the former case, numerical noise in the derivatives
of $E(z)$ create certain scatter in $w_{DE}(z)$ at late times.

\end{document}